\documentclass[%
 reprint,
 amsmath,amssymb,
 aps,
floatfix,
]{revtex4-2}

\usepackage{graphicx}% Include figure files
\usepackage{dcolumn}% Align table columns on decimal point
\usepackage{bm}% bold math
\usepackage[figuresright]{rotating}
\usepackage{url}
\usepackage{siunitx}
\usepackage{cleveref}
\crefname{equation}{Eq}{Eqs}
\crefrangelabelformat{equation}{(#3#1#4--#5#2#6)}

\usepackage{breakcites}
\begin{document}

\preprint{APS/123-QED}

\title{Sympletic tracking methods for insertion devices:\\ a Robinson wiggler example}% Force line breaks with \\
%%\thanks{A footnote to the article title}%

% \altaffiliation[Also at ]{Physics Department, XYZ University.}%Lines break %automatically or can be forced with \\
\author{Ji Li}%
 \email{ji.li@helmholtz-berlin.de}
\affiliation{%
  Helmholtz-Zentrum Berlin für Materialien und Energie GmbH (HZB), Albert-Einstein-Straße 15, 12489 Berlin, Germany%%\\This line break forced with \textbackslash\textbackslash
}%
\author{Jörg Feikes}%
\affiliation{%
  Helmholtz-Zentrum Berlin für Materialien und Energie GmbH (HZB),  Albert-Einstein-Straße 15, 12489 Berlin, Germany}%
  
\author{Tom Mertens}
\affiliation{%
  Helmholtz-Zentrum Berlin für Materialien und Energie GmbH (HZB), Albert-Einstein-Straße 15, 12489 Berlin, Germany}%

\author{Edward Rial}%
\affiliation{%
  Helmholtz-Zentrum Berlin für Materialien und Energie GmbH (HZB),  Albert-Einstein-Straße 15, 12489 Berlin, Germany}%
\author{Markus Ries}%
\affiliation{%
  Helmholtz-Zentrum Berlin für Materialien und Energie GmbH (HZB),  Albert-Einstein-Straße 15, 12489 Berlin, Germany}%
\author{Andreas Schälicke}%
\affiliation{%
  Helmholtz-Zentrum Berlin für Materialien und Energie GmbH (HZB),  Albert-Einstein-Straße 15, 12489 Berlin, Germany}%
  
\author{Luis Vera Ramirez}
\affiliation{%
  Helmholtz-Zentrum Berlin für Materialien und Energie GmbH (HZB),  Albert-Einstein-Straße 15, 12489 Berlin, Germany}%

\date{\today}% It is always \today, today,

\begin{abstract}
Modern synchrotron light sources are often characterized with high-brightness synchrotron radiation from insertion devices. Inevitably, insertion devices introduce nonlinear distortion to the beam motion. Symplectic tracking is crucial to study the impact, especially for the low- and medium-energy  storage rings. This paper uses a Robinson wiggler as an example to illustrate an universally applicable analytical representation of the magnetic field and to summarizes four different symplectic tracking methods.
\end{abstract}

\maketitle

\section{introduction}

With the aim of high-brightness synchrotron radiation, the storage rings of modern synchrotron light sources mostly adopt strong-focusing lattices, which result in large negative natural chromaticities and need strong sextupoles to correct the chromaticity to suppress the head-tail instability. Therefore nonlinear distortion is introduced to beam motion by strong sextupole fields. Furthermore, insertion devices, fringe fields and imperfections of magnets are additional sources of nonlinearity. The nonlinear distortion from the magnets determines long-term beam stability and has strong impact on operational performance.

The analysis of long-term beam dynamics in the storage ring is established by symplectic particle tracking. In general, symplectic tracking can be divided into two steps. First, an accurate analytical expression of magnetic field is needed. Second, the symplectic integration to solve the Hamiltonian equations of the particle's motion inside the magnetic field is conducted stepwise element by element for multiple turns. Unlike the  Runge-Kutta integration which is usually not sympletic and may introduce artificial damping and antidamping effect, sympletic integration leads to the canonical transformation of phase space vector and satisfies Liouville's theorem.

In tracking codes the effect of dipoles and multipoles are usually modeled with an impulse boundary approximation, also called hard-edge model, in which the magnetic field is assumed to be constant within the effective boundary of the magnet and zero outside. In this model, only the longitudinal component of the vector potential is needed to describe the system. Since the coordinates and their conjugate canonical momenta are not mixed in the Hamiltonian, the Hamiltonian can be split into drift-kick combinations~\cite{yoshida_1990}. 

\begin{figure}[hbt!]
   \setlength\belowcaptionskip{-2.0\baselineskip}
   \centering
    \includegraphics*[width=0.65\columnwidth]{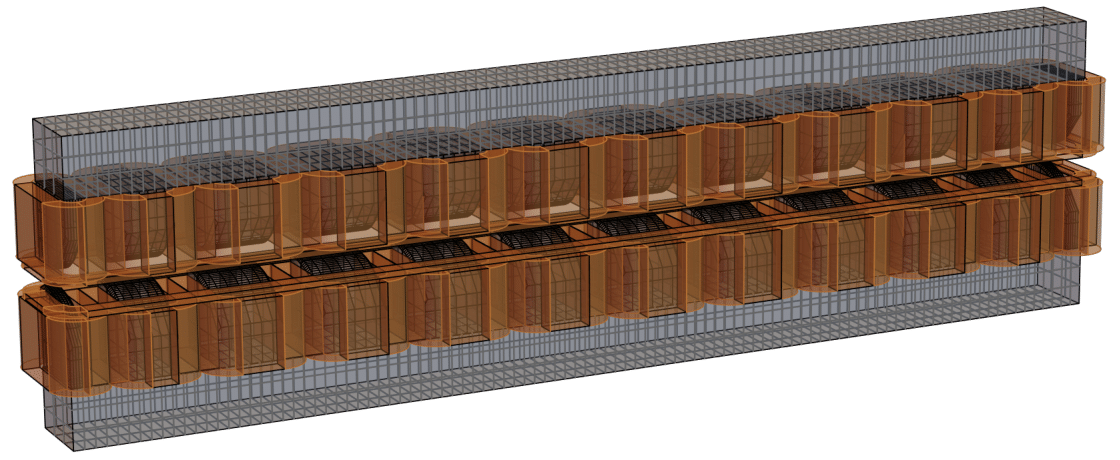}
    \caption{ The model of the RW in RADIA~\cite{chubar_elleaume_chavanne_1998}.}
    \label{RWmodel}
\end{figure}

\begin{figure}[hbt!]
   \centering
    \includegraphics*[width=0.75\columnwidth]{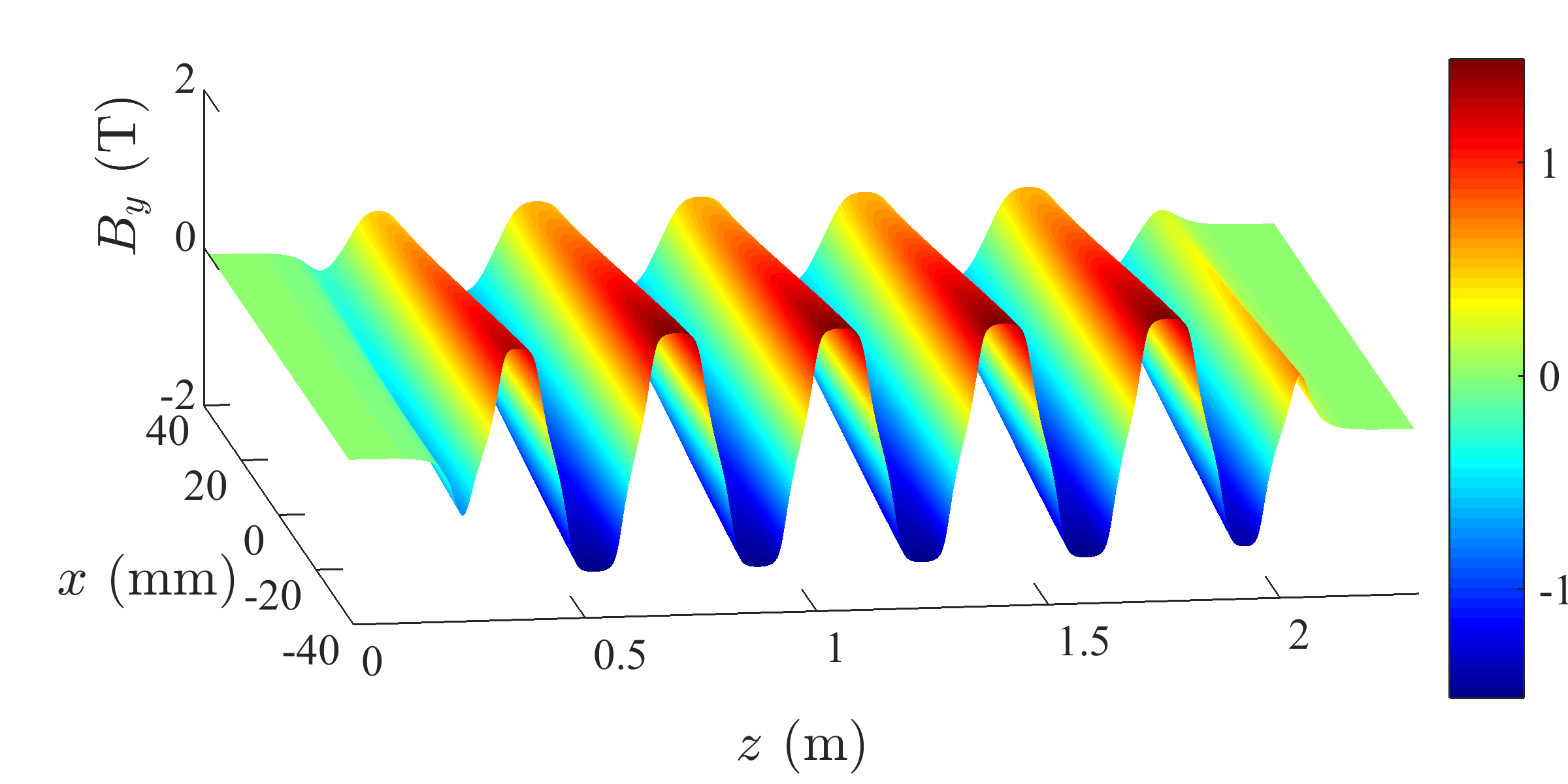}
    \caption{The vertical magnetic field on the midplane of the RW.}
    \label{RWfield}
\end{figure}

The proposed Robinson Wiggler (RW) for the Metrology Light Source (MLS)~\cite{MLSgeneral}, designed and studied in Ref.~\cite{Tydecks2016A}, is used to illustrate symplectic tracking methods for insertion devices. It consists of a chain of 12 combined-function magnets, shown in Fig.~\ref{RWmodel}, with the aim to lengthen the bunch by transferring the longitudinal damping to transverse plane. As shown in Fig.~\ref{RWfield}, the magnetic field in the RW is three-dimensional (3D), horizontally asymmetric and much more complicated than the impulse boundary model,  thus the splitting methods for dipoles and multipoles are not applicable any more.

In this paper, the principle of the RW and the necessity of symplectic tracking is briefly introduced in section~\ref{motivation}. Then in section~\ref{basicconcepts} the basic concepts for symplectic integration are revisited. In section~\ref{Afield} an analytical representation is proposed to describe the 3D field in the RW accurately. On this basis, three sympletic integration methods are introduced to solve the Hamiltonian equations of motion for electrons in section~\ref{integratorswithAfield}. In section~\ref{trackingmono}, a monomial map approach independent of analytic expression of the magnetic field is introduced to realize faster tracking. The methods in this paper are universally applicable to all wigglers and undulators with a straight reference trajectory.

\begin{figure*}[t!]
   \centering
   \includegraphics*[width=\textwidth]{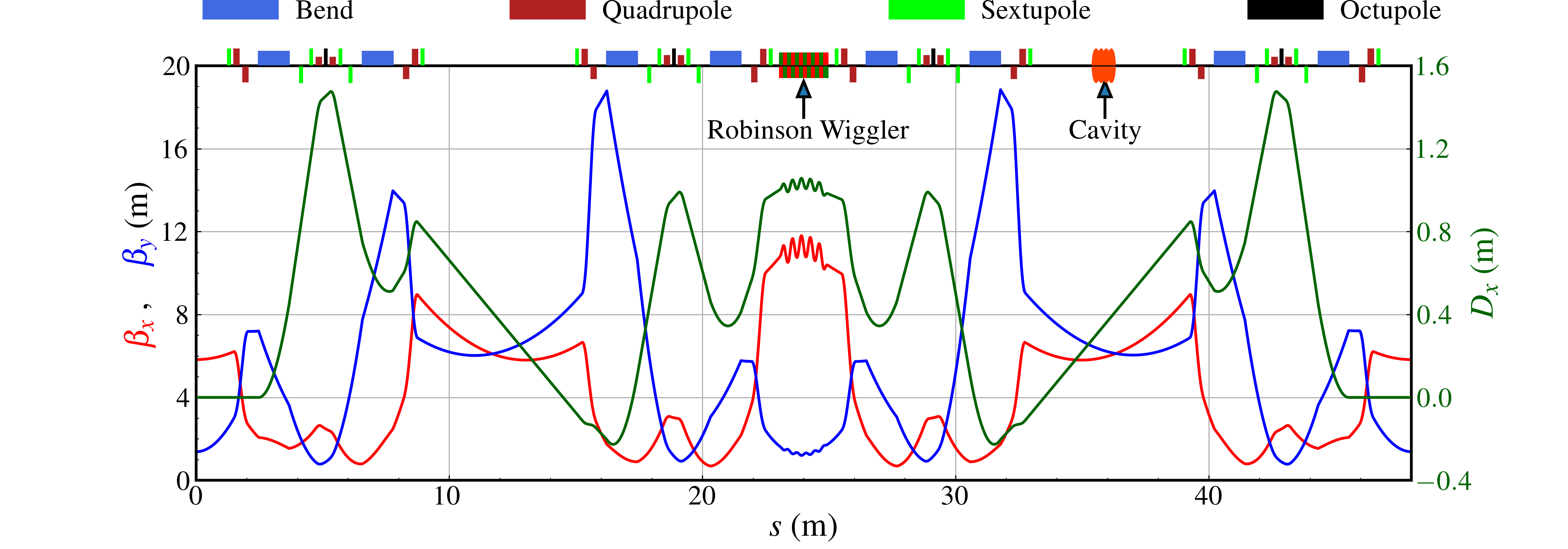}
     \caption{Linear optics of the MLS with Robinson wiggler.}
   \label{linearoptics}
\end{figure*}

\section{Motivation: a Robinson Wiggler for the Metrology Light Source}\label{motivation}

The Metrology Light Source (MLS) is an electron storage ring owned by the Physikalisch-Technische Bundesanstalt (PTB) and operated and
designed by the Helmholtz-Zentrum Berlin für Materialien und Energie (HZB). It
is dedicated to metrology applications in the Ultraviolet (UV)
and Extreme violet (EUV) spectral range as well as in the Infrared (IR)  and THz region~\cite{feikes2011}. It can be
operated at any energy between 50 MeV and 629 MeV, while the stored current can
be varied from 200 mA down to a single electron (= 1 pA). The main parameters of the major operational mode, standard user mode, at the MLS are listed in Table~\ref{MLSpara}.

\begin{table}[hbt!]
\caption{Parameters of the standard user mode at the MLS}
\label{MLSpara}
\begin{ruledtabular}
\begin{tabular}{ll}
\textrm{Parameter}&
\textrm{Value}\\
\colrule
Operation Energy & 629 MeV\\
Injection energy & 105 MeV \\
Tunable energy range & 50 - 629 MeV \\
Tunable current range & 1 pA - 200 mA \\
Circumference & 48 m \\
Horizontal/vertical tunes & 3.178 / 2.232 \\
Short/long straight & 2.5 m / 6 m \\
Natural emittance & 110 nm rad @ 629 MeV\\
Natural energy & 4.4 $\times$ $10^{-4}$ @ 629 MeV\\
Momentum compaction factor & 0.03 \\
Lifetime @ 150 mA, 629 MeV & $\sim$ 6 h \\
\end{tabular}
\end{ruledtabular}
\end{table}

\begin{figure}[hbt!]
   \centering
   \includegraphics*[width=\columnwidth]{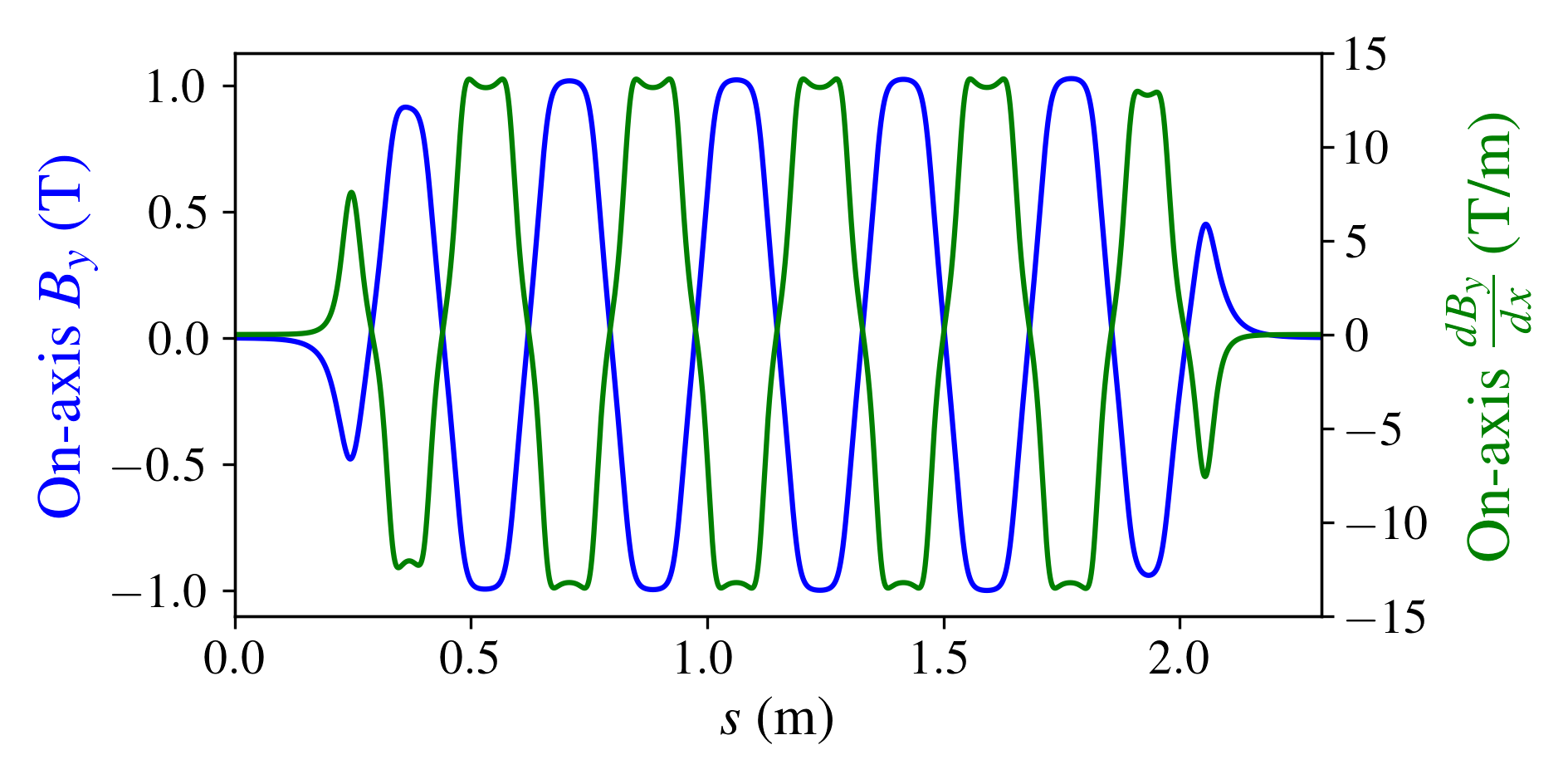}
     \caption{On-axis dipole and quadrupole components of the RW.}
   \label{fieldcomp}
\end{figure}

The MLS is operated in decay mode. The standard user mode has a beam lifetime of $\sim$ 6 hours at 150 mA and therefor requires 2-3 injections per day. Each injection interrupts the user operation for approximately  30 minutes and affects the users' experiments for another nearly 1 hour due to thermal load changes on the components of optical beamlines after the injection. Therefore a RW, a chain of combined function magnets, was proposed to be installed in the dispersive straight section in the storage ring of the MLS to increase the beam lifetime, noted in the Fig.~\ref{linearoptics}. The major parameters are listed in Table~\ref{RWpara}. 
%The optimized hyperbolic pole shape results in a combination of vertical magnetic field, and its gradient, on the beam path, shown in Fig.~\ref{fieldcomp}.

\begin{table}[hbt!]
\caption{Parameters of the RW}
\label{RWpara}
\begin{ruledtabular}
\begin{tabular}{ll}
\textrm{Parameter}&
\textrm{Value}\\
\colrule
wiggler length & 1.9 m\\
number of  poles & 12 \\
central pole length & ~110.47 mm \\
end pole length &  ~82.85 / 27.62 mm \\
period length & ~354.78 mm \\
maxium on-axis By & 1 T\\
\end{tabular}
\end{ruledtabular}
\end{table}

According to ~\cref{I2,I4,D,emittance,bunchlength}, the vertical magnetic field and its gradient inside the RW shown in Fig.~\ref{fieldcomp} together with the positive dispersion yields a negative value of $I_4$, thus negative damping partition $D$. Therefore the transverse emittance $\epsilon_x$ can be reduced by transferring longitudinal damping to the horizontal plane, while the bunch is lengthened due to increased energy spread $\sigma_{\delta}$ ~\cite{Tydecks2016A}. With the vertical white noise excitation acting on the beam to keep the transverse beam size the same as that in standard user mode,  the lifetime is increased to $\sim$ 12~hours at 150~mA because of the increased bunch volume.

\begin{equation}\label{I2}
\begin{aligned}
I_2  = \oint \frac{1}{\rho^2} ds ,
\end{aligned}
\end{equation}

\begin{equation}\label{I4}
\begin{aligned}
I_4  = \oint (\frac{\eta_x}{\rho}+ 2\eta_x\frac{B_y}{B\rho}\frac{1}{B\rho}\frac{\partial B_y}{\partial x} ) ds ,
\end{aligned}
\end{equation}

\begin{equation}\label{D}
\begin{aligned}
D = \frac{I_4}{I_2},
\end{aligned}
\end{equation}

\begin{equation}\label{emittance}
\begin{aligned}
\epsilon_x \propto \frac{1}{1-D},
\end{aligned}
\end{equation}

\begin{equation}\label{bunchlength}
\begin{aligned}
\sigma_{\delta} \propto \frac{1}{1+D}.
\end{aligned}
\end{equation}

The maximum on-axis $B_y$ ($\sim$ \SI{1}{T}) is close to the dipole strength($\sim$ \SI{1.373}{T}) in the bending magnet. Although the RW was carefully designed and optimized, the nonlinear distortion of this strong and long-period ($\sim$0.355 m for one period) insertion device to the stored beam in the low-energy storage ring is of concern and should be verified with symplectic tracking.

\section{basic concepts for symplectic tracking}\label{basicconcepts}
The problem studied in this paper is the motion of a particle moving through a static magnetic field with a straight reference trajectory. The magnetic field is described by a vector potential $A = (A_x, A_y , A_z)$ in Cartesian coordinate system, so the Hamiltonian for the motion of a particle is:
\begin{eqnarray}\label{Hamiltionian_acc}
\begin{aligned}
& H = \frac{\delta}{\beta_0} - a_z  - \\
&\sqrt{(\frac{1}{\beta_0} + \delta)^2 - (p_x-a_x)^2 - (p_y-a_y)^2- \frac{1}{\beta_0^2 \gamma_0^2}}.  
\end{aligned}
\end{eqnarray}
where a particle with charge $q$  and the reference momentum $P_0$ has velocity $\beta_0c$ and relativistic factor $\gamma_0 = (1 - {\beta_0}^2)^{-{\frac{1}{2}}}$
and the scaled vector potential $a=(a_x, a_y , a_z) = q(A_x, A_y , A_z)/P_0$. 

The dynamical variables used in beam dynamics are defined in the following way: the horizontal and vertical transverse coordinates are $x$ and $y$, respectively; their corresponding momenta $p_x$ and $p_y$ are defined as:

\begin{eqnarray}
p_x = \frac{\gamma m\dot{x}+qA_x}{P_0},
\end{eqnarray}
\begin{eqnarray}
p_y = \frac{\gamma m\dot{y}+qA_y}{P_0}.
\end{eqnarray}

The longitudinal coordinate is usually expressed as $z$, however,  $l$ is used to be distinguished from the physical meaning of subscript $z$ in Eq.~\eqref{Hamiltionian_acc}. 
\begin{equation}\label{ldefinition}
\begin{aligned}
l = \frac{s}{\beta_0} - ct.
\end{aligned}
\end{equation}
where the particle arrives at position s along the reference trajectory at time t assuming s = 0 at time t = 0 for the reference particle.

The longitudinal momentum, referred to as the energy deviation, is written:
\begin{equation}\label{deltadefinition}
\begin{aligned}
\delta = \frac{E}{cP_0} - \frac{1}{\beta_0}.
\end{aligned}
\end{equation}

The three pairs of canonical variables $(x,p_x)$, $(y,p_y)$, $(l,\delta)$ should satisfy the Hamiltonian equations Eq.~\eqref{H-eq1} and Eq.~\eqref{H-eq2}~\cite{wolski_2014}. 

\begin{eqnarray}\label{H-eq1}
    \frac{d q_i}{d s}  = \frac{\partial H}{\partial p_i},
\end{eqnarray}
\begin{eqnarray}\label{H-eq2}
    \frac{d p_i}{d s}  = -\frac{\partial H}{\partial q_i}.
\end{eqnarray}
where $q_i=x$, $y$, $l$ and $p_i=p_x$, $p_y$, $\delta$, respectively.

The transformation of the particle from the one position $s$ to the next $s+\Delta s$, equivalent to the solutions of Eq.~\eqref{H-eq1} and Eq.~\eqref{H-eq2}, can be represented by a transfer map  $\mathcal{M}$ in the six-dimensional phase space of the canonical coordinates of the particle:

%\begin{equation}\label{transfermap}
%\begin{aligned}
%&\vec{X} = (x,p_x,y,p_y,z,\delta)\big|_{s+\Delta s}\\
%&\vec{x} = (x,p_x,y,p_y,z,\delta)\big|_{s}\\
%&\vec{X}= \mathcal{M}\vec{x}
%\end{aligned}
%\end{equation}
\begin{subequations}
\label{transfermap}
\begin{eqnarray}
\begin{aligned}
&\vec{X} = (x,p_x,y,p_y,z,\delta)\big|_{s+\Delta s} ,\label{transfermap_a}
\end{aligned}
\end{eqnarray}

\begin{equation}
\begin{aligned}
&\vec{x} = (x,p_x,y,p_y,z,\delta)\big|_{s},\label{transfermap_b}
\end{aligned}
\end{equation}
\begin{equation}
\begin{aligned}
&\vec{X}= \mathcal{M}\vec{x}.\label{transfermap_c}
\end{aligned}
\end{equation}
\end{subequations}

It is important that transformation preserves the symplectic nature of the dynamics, otherwise use of non-symplectic transfer maps can lead to artificial growth or damping of the beam motion, resulting in inaccurate information on the long-term stability of the beam motion. The criterion of symplectic transformation is: 

\begin{equation}\label{symplecticity}
\begin{aligned}
\emph{J}^T \cdot  \emph{S}  \cdot  \emph{J} = \emph{S}.
\end{aligned}
\end{equation}
where the $\emph{J}$ is the Jacobian of the transformation from $s$ to $s+\Delta s$,

\begin{equation}\label{Jacobian}
J_{ij} = \frac{\partial {X_i}}{\partial {x_j}}.
\end{equation}
and $\emph{S}$ is a block-diagonal matrix constructed from $2\times2$ antisymmetric matrices $S_2$ :

\begin{equation}\label{Smatrix}
\begin{aligned}
   S_2  = \begin{pmatrix}
  0   & 1 \\
  -1 &   0
  \end{pmatrix}.
\end{aligned}
\end{equation}

Above all, the core content of tracking particles through insertion devices is symplectic integration of the Hamiltonian equations Eq.~\eqref{H-eq1} and Eq.~\eqref{H-eq2}. Obviously, the derivatives of the vector potentials are needed, therefore an accurate analytic representation of the magnetic field is key.  

\section{Analytical representation of the magnetic field in the Robinson wiggler}\label{Afield}
Usually we have the measured or numerical 3D magnetic field data on a discrete mesh of points throughout the region of interest. However, the discrete field map cannot be directly used for symplectic tracking and should be described by analytical formulae. Various representations have been included in Ref.~\cite{wu_forest_robin_2003,bahrdt_wuestefeld_2011,giboudot_wolski_2012,titze_bahrdt_wuestefeld_2016}, especially Mitchell has done systematic work in Ref.~\cite{mitchell2007} on describing the magnetic field with generalized gradient in different coordinate systems. In this paper, we stick to the Halbach expression in a Cartesian coordinate system.

The Halbach expansions of the magnetic field in planer undulators or wigglers can be expressed in the following ~\cite{wu_forest_robin_2003,wolski_2014}, which satisfy Maxwell's equations and Laplace's equation.

\begin{equation}\label{Hbx}
 B_x= -\sum_{m,n}^{M,N}\frac{C_{mn}mk_x}{k_{y,mn}}sin(mk_xx)sinh(k_{y,mn}y)sin(nk_{z}z),
\end{equation}

\begin{equation}\label{Hby}
 B_y= \sum_{m,n}^{M,N}C_{mn}cos(mk_xx)cosh(k_{y,mn}y)sin(nk_{z}z),
\end{equation}
\begin{equation}\label{Hbz}
 B_z= \sum_{m,n}^{M,N}C_{mn}\frac{nk_z}{k_{y,mn}}cos(mk_xx)sinh(k_{y,mn}y)cos(nk_{z}z),
\end{equation}
\begin{equation}\label{Hky}
   k_{y,mn}^{2} = m^2k_x^2+n^2k_z^2.
\end{equation}
in which M and N represent the maximum numbers of harmonics in $x$ and $z$ directions.

As depicted in Fig.~\ref{RWfield}, the vertical magnetic field $B_y$ on the midplane is horizontally asymmetric, which cannot be described with Eq.~\eqref{Hby}. It is necessary to modify the Halbach expansions by adding $\theta$ in $cos$ terms. In practice, faster convergence is gained by adding $\theta_{mn}$ and $\phi_{mn}$ in the $cos$ and $sin$ terms in Eq.~\eqref{Hby}. Therefore Equation.~\eqref{Hby} is modified to the new form shown in Eq.~\eqref{Mby}.

\begin{equation}\label{Mby}
\begin{aligned}
    B_y = \sum_{m,n}^{M,N}&C_{mn}cos(mk_xx+\theta_{mn})cosh(k_{y,mn}y)\\
                          &\times sin(nk_{z}z +\phi_{mn})
\end{aligned}
\end{equation}

\begin{figure*}[t!]
   \centering

   \includegraphics*[width=\textwidth]{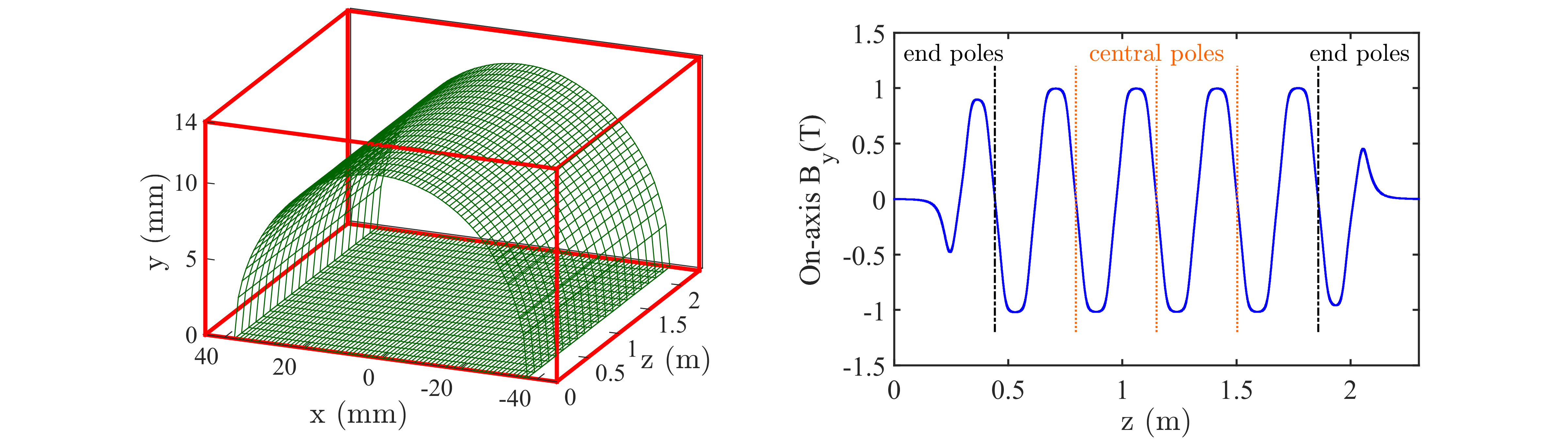}
     \caption{Fitting approach: (left). The region of the field map and the geometry of vacuum chamber.   (right). Splitting of the field map for Fourier decomposition. In the left plot the field map for Fourier decomposition is bounded with red frame, and the geometry of the vacuum chamber is marked as green meshes. The right plot shows that there are four central periods, consisting of eight central poles;and two end periods but four end poles.}
   \label{fitting}
\end{figure*}

The 3D gridded field map, bounded by the red frame in the left plot of Fig.~\ref{fitting}, is used for Fourier decomposition (field fitting). It covers the range from -40 to 40~$mm$ horizontally, from 0 to 14~$mm$ vertically and from 0 to 2.3~$m$ longitudinally. And the grid size of the field map is 1 mm in transverse plan and 5 mm in longitudinal direction. Due to the symmetry in vertical direction, the region of $y < 0$ is not displayed in the left plot of Fig.~\ref{fitting}. Considering the complexity of the field, it takes too many coefficients to apply Fourier decomposition based on Eq.~\eqref{Mby}  to the whole Robinson wiggler which results in very slow convergence of the fit. Instead the whole field map of the RW should be divided into two end-pole sections and one central-pole section, as marked in the right plot of Fig.~\ref{fitting}. In principle, the entrance field of the end poles can be treated identical with the exit one after coordinate transformation, and the field of each period in the central section can be considered identical as well. Therefore, the Fourier decomposition is only needed for the entrance section and one period in the central section. 

%\begin{figure}[hbt!]
%   \centering
%   \includegraphics*[width=\columnwidth]{fittingzone5}%fitting_method
%     \caption{Fitting approach: (top). The region of the field map and the geometry of vacuum chamber.   (bottom). Splitting of the field map for Fourier decomposition. In the top plot the field map for Fourier decomposition is bounded with red frame, and the geometry of the vacuum chamber is marked as green meshes. The bottom plot shows that there are four central periods, consisting of eight central poles;and two end periods but four end poles}
%   \label{fitting}
%\end{figure}

The accuracy of the Fourier decomposition in the Region of Interest (ROI),  bounded by the vacuum chamber geometry, is crucial for the nonlinear beam dynamics simulation. The horizontal and vertical apertures of the vacuum chamber are $\pm$ 37.5~$mm$ and $\pm$ 12~$mm$, respectively. Only the upper half of the ROI is noted in Fig.~\ref{fitting} left plot due to the vertical symmetry, which is enclosed by the midplane and upper half elliptical vacuum chamber geometry, marked as green meshes. 

\begin{widetext}
\begin{figure*}[htb]
\centering
\begin{minipage}[t]{86mm}
\centering
\includegraphics[width=86mm]{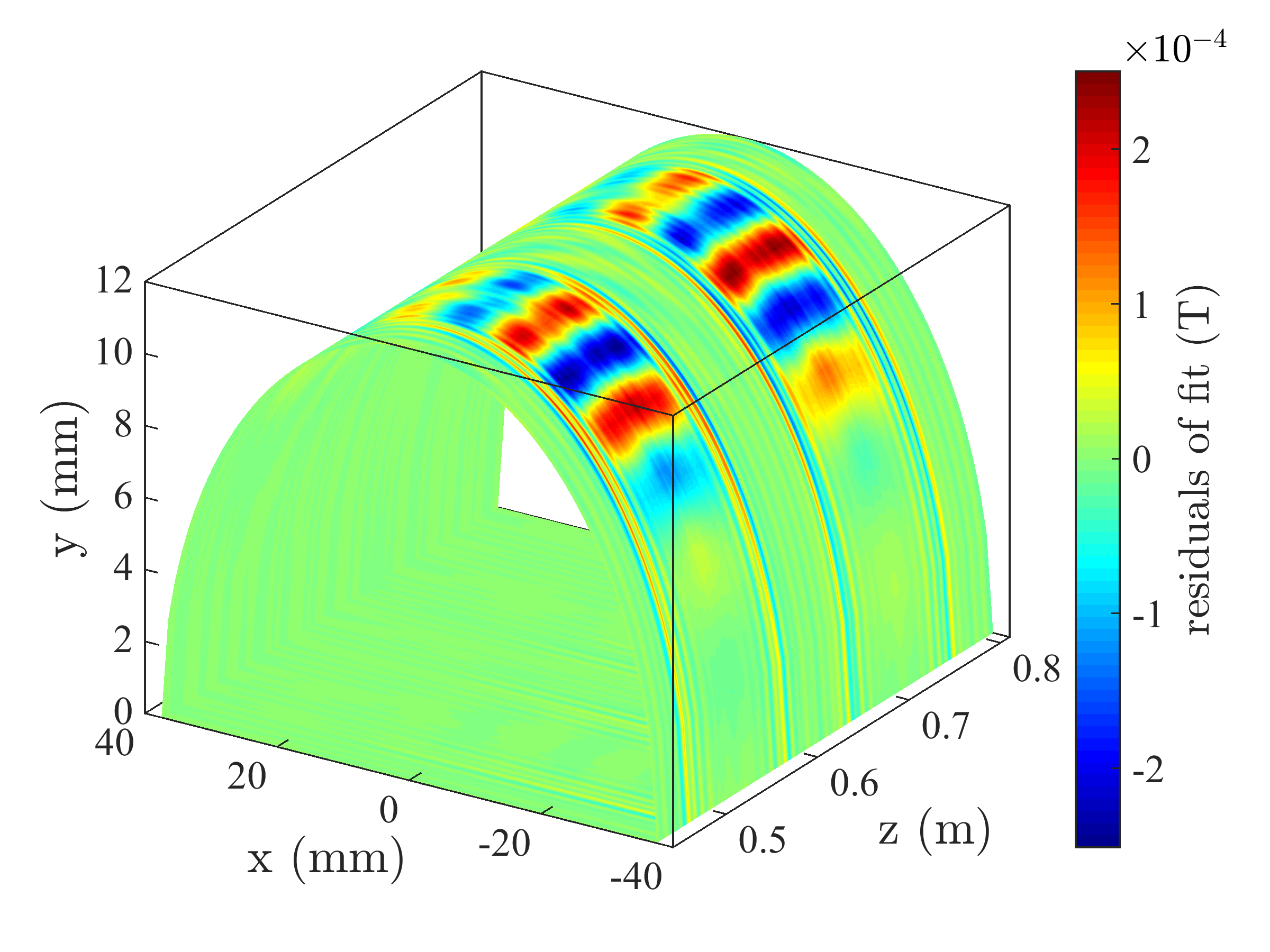}
\caption{Comparison of the numerical field map in the first period of the main poles from RADIA and its analytic expression.}
\label{residual2}
\end{minipage}\hfill
\begin{minipage}[t]{86mm}
\centering
\includegraphics[width=86mm]{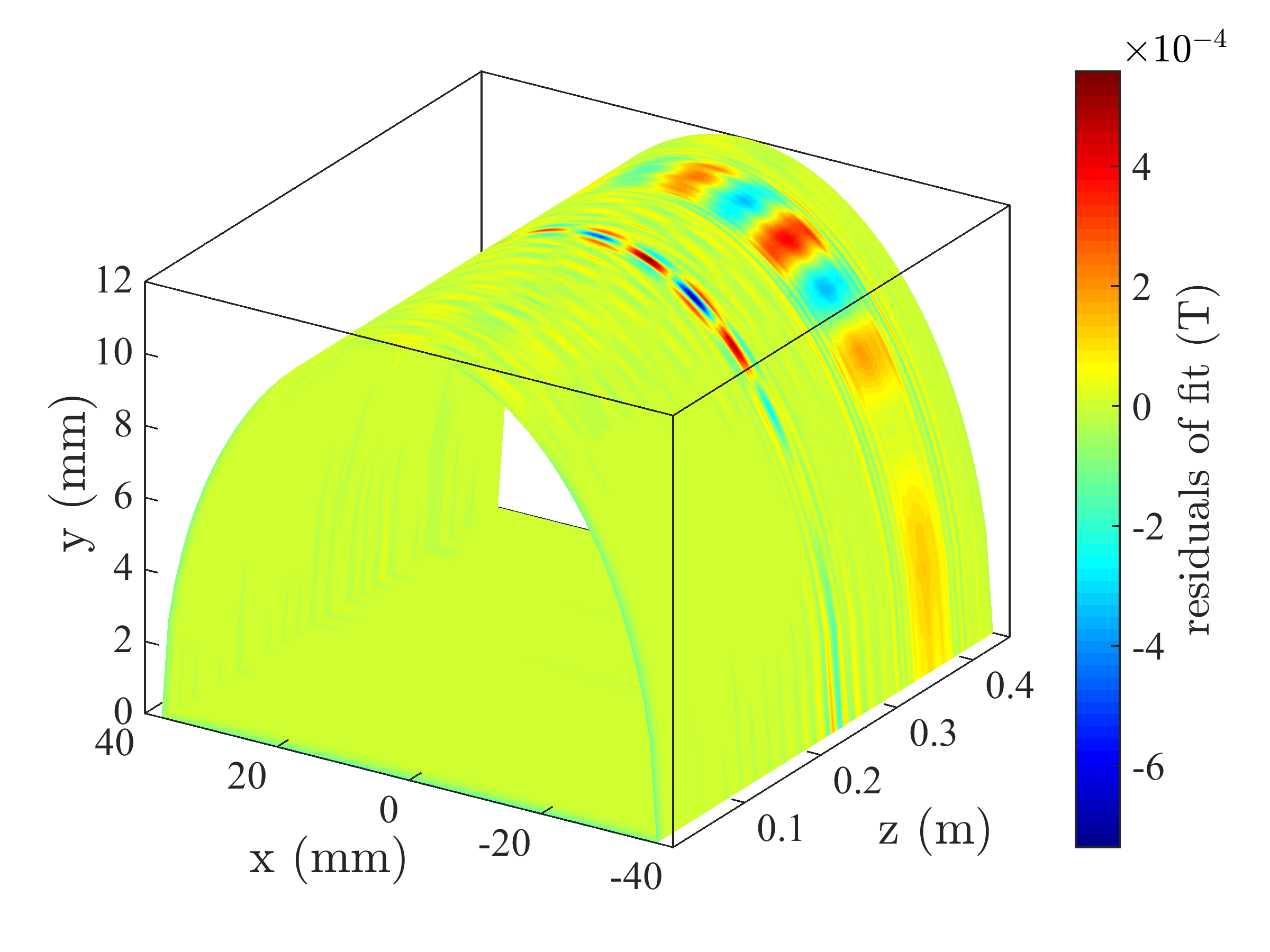}
\caption{Comparison of the numerical field map in the entrance end-pole section from RADIA and its analytic expression.}
\label{residual1}
\end{minipage}
\end{figure*}
\end{widetext}

Based on Eq.~\eqref{Mby}, the coefficents $C_{mn}$,  $\theta_{mn}$ and $\phi_{mn}$  can be fitted to numerical 3D field map from RADIA~\cite{chubar_elleaume_chavanne_1998}. As shown in Fig.~\ref{residual2}, the maximum residual of the analytical field representation with M =20 and N=65, on the upper surface of the vacuum chamber in the central-pole section, is $\sim$ $2.5 \times 10^{-4}$~T, even below $\sim$ $7 \times 10^{-5}$~T on the midplane. With $cosh$ terms in Eq.~\eqref{Mby}, the residuals increase exponentially with $y$, which means the residuals in the region of interest are below $\sim$ $2.5 \times 10^{-4}$~T. Similarly shown in Fig.~\ref{residual1}, the residuals in the end-pole section at the entrance are blow $7 \times 10^{-4}$~T with M=20 and N=85. Above all, the modified Halbach expressions can describe the magnetic field in the RW accurately enough for sympletic tracking. In a broader sense, Equation.~\eqref{Mby} is universally applicable to undulators and wigglers with vertically symmetrical field, which describes a large range of the insertion devices. In addition, further modified expressions based on Eq.~\eqref{Mby} for an APPLE II udulator are given and verified in Appendix~\ref{APPLIIappendix}.

Reconstructing the 3D magnetic field analytically from discrete field maps is in general very challenging. The fitting routine deals with thousands of coefficients, and uses parallel computation in Python~\cite{Python} together with Cython~\cite{Cython} and Intel Math Kernel Library~\cite{MKL_intel}  to achieve the desired accuracy within reasonable time budget(6-8 weeks). In the late phase of this work, CUDA GPU acceleration in Pytorch~\cite{Pytorch} is used as alternative fitting approach. Nevertheless, preparing the analytical representation is the most time-consuming part for symplectic tracking.

\section{symplectic integration based on analytical field representation}\label{integratorswithAfield}
Tracking particles over multiple turns in the MLS storage ring, realized by symplectic integration of Hamiltonian equations of motion, is an major approach to study the nonlinear distortion of the RW on the beam motion. Therefore an analytical form of vector potential is needed. When the analytical representation of the vertical magnetic field is established, the scalar potential can be derived as Eq.~\eqref{scalarpotential} shows. Accordingly the vector potential $A_x$ and $A_y$ can be expressed as Eq.~\eqref{vectorpotentialx} and Eq.~\eqref{vectorpotentialy} with the chosen gauge~$A_{z}=0$.
\begin{widetext}
\begin{equation}
 V = -\int B_y dy = -\sum_{m,n}\frac{c_{m,n}}{k_{y,mn}}cos(mk_xx+\theta_{m,n})sinh(k_{y,mn}y)sin(nk_{z}z+ \phi_{m,n})\;, 
\label{scalarpotential}
\end{equation}

\begin{equation}
A_x =- \int_{z_0}^{z} \frac{\partial V}{\partial y} dz + C_1 = -\sum_{m,n}c_{m,n}\frac{1}{nk_z}cos(mk_xx+\theta_{m,n})cosh(k_{y,mn}y)cos(nk_{z}z + \phi_{m,n})\;, 
\label{vectorpotentialx}
\end{equation}

\begin{equation}
A_y =- \int_{z_0}^{z} \frac{\partial V}{\partial x} dz + C_2 = -\sum_{m,n}\frac{c_{m,n}mk_x}{k_{y,mn}nk_z}sin(mk_xx+\theta_{m,n})sinh(k_{y,mn}y)cos(nk_{z}z + \phi_{m,n})\;,
\label{vectorpotentialy}
\end{equation}

\begin{equation}
A_z = 0\;. 
\label{vectorpotentialz}
\end{equation}

\end{widetext}

In the following section three different sympletic integrators will be introduced. Multi-turn tracking is conducted with ELEGANT. Tracking though the RW is accomplished through the SCRIPT element in ELEGANT to call the customized symplectic integrators and return the particle coordinates to ELEGANT. It is worth noting that in general ELEGANT uses $(x,x',y,y', s, dP/P_0)$ to describe the motion of a particle. Necessary conventions in~\crefrange{cordsconversion_xp}{cordsconversion_delta}are included in customized integrators, which are implemented with Python and Fortran.

\begin{equation}\label{cordsconversion_xp}
 x^{\prime} = \frac{p_x-a_x}{\sqrt{(\frac{1}{\beta_0} + \delta)^2 - (p_x-a_x)^2 - (p_y-a_y)^2- \frac{1}{\beta_0^2 \gamma_0^2}}},
\end{equation}

\begin{equation}\label{cordsconversion_yp}
 y^{\prime} = \frac{p_y-a_y}{\sqrt{(\frac{1}{\beta_0} + \delta)^2 - (p_x-a_x)^2 - (p_y-a_y)^2- \frac{1}{\beta_0^2 \gamma_0^2}}}, 
\end{equation}

\begin{equation}\label{cordsconversion_delta}
   \delta = \frac{P}{P_0\beta}- \frac{1}{\beta_0},
\end{equation}

\subsection{Implicit Runge-Kutta integrator} 
The Runge-Kutta method can be used to integrate the Hamiltonian equations of motion, however, the integration will only be symplectic for specific Butcher tableaux~\cite{butcher_2016}. Applying the implicit-midpoint integrator~\cite{wolski_2014}, a second order Runge-Kutta integrator, the Eq.~\eqref{H-eq1} and Eq.~\eqref{H-eq2} can be rewritten as Eq.~\eqref{irkx} and Eq.~\eqref{irkpx}:
\begin{equation}\label{irkx}
    x(s+\Delta s) = \left. x(s)+\Delta s \frac{\partial H}{\partial p_x}\right|_{x=x_m^{(1)},p_x=p_{xm}^{(1)}},
\end{equation}

\begin{equation}\label{irkpx}
    p_x(s+\Delta s) = \left. p_x(s)-\Delta s \frac{\partial H}{\partial x}\right|_{x=x_m^{(1)},p_x=p_{xm}^{(1)}}.  
\end{equation}
in which the intermediate values $x_m^{(1)}$ and $p_{xm}^{(1)}$ can be solved from Eq.~\eqref{irkxmid} and Eq.~\eqref{irkpxmid} with the Newton-Raphson method~\cite{numerical_recipe2020}.

\begin{equation}\label{irkxmid}
    x_m^{(1)} = \left. x(s)+\frac{1}{2} \Delta s \frac{\partial H}{\partial p_x}\right|_{x=x_m^{(1)},p_x=p_{xm}^{(1)}},
\end{equation}

\begin{equation}\label{irkpxmid}
    p_{xm}^{(1)} = \left. p_x(s)-\frac{1}{2} \Delta s \frac{\partial H}{\partial x}\right|_{x=x_m^{(1)},p_x=p_{xm}^{(1)}}.  
\end{equation}

It can be seen from the above, that the implicit midpoint integration is straightforward and easy to implement. However, the Newton-Raphson method is applied to each step of the integration to solve a set of algebraic equations, therefore the computational cost is rather expensive.

\subsection{Wu-Forest-Robin integrator} 

Wu, Forest and Robin developed an explicit symplectic integrator for the charged particle Hamiltonian with an s-dependent static magnetic field~\cite{wu_forest_robin_2003,wolski_2014}. The basis of this method is to extend phase space by making $z$ a dynamic variable, introducing  a new canonical momentum $p_{z}$ conjugate to $s$, as well as a new independent variable $\sigma$. The equivalent Hamiltonian in the extended phase space is given by

\begin{equation}\label{newH}
\overline{H} = H + p_z.
\end{equation}
The equations of motion for the new dynamics variables ($s$,$p_{z}$) are:
\begin{equation}\label{eq_H_z}
\begin{aligned}
\frac{d z}{d \sigma}  = \frac{\partial {\overline{H}} }{\partial p_z} = 1,
\end{aligned}
\end{equation}
\begin{equation}\label{eq_H_pz}
\begin{aligned}
\frac{d p_z}{d \sigma}  = -\frac{\partial {\overline{H}}}{\partial z}= -\frac{\partial H}{\partial z}.
\end{aligned}
\end{equation}

To simplify the integration, the old and new independent variables are expressed as:
\begin{equation}\label{zsigma}
\begin{aligned}
z = \sigma.
\end{aligned}
\end{equation}

The particle motion is now described by $x$, $y$, $l$ and $z$, together with their corresponding canonical momenta. The Hamiltonian of the Robinson wiggler in the extended phase space has no explicit dependence on $\sigma$, the evolution of function f (which represents any of the dynamic variables) can be expressed in terms of a Lie transformation:
\begin{equation}\label{lietrans}
    \left. f \right|_{\sigma=\sigma_{0}+\Delta \sigma} = \left. e^{-\Delta \sigma:\overline{H}:}f\right|_{\sigma=\sigma_{0}}.
\end{equation}

Now the Hamiltonian in extended phase space is:

\begin{equation}
\begin{aligned}
& \overline{H} = \frac{\delta}{\beta_0} \\
& -\sqrt{(\frac{1}{\beta_0} + \delta)^2 - (p_x - a_x)^2 - (p_y-a_y)^2- \frac{1}{\beta_0^2 \gamma_0^2}} +p_z\label{WFRHamiltonian}.
\end{aligned}
\end{equation}

In order to use Baker-Campbell-Hausdorff and Zassenhaus formulae~\cite{Dragt2020} to split the Hamiltonian into integrable terms, a paraxial approximation is made by
expanding the square root to the second order~\cite{wolski_2014}. Thus $\overline{H}$ is split into $H_1$, $H_2$, $H_3$.

\begin{equation}\label{H1H2H3}
\overline{H} \approx H_{1} + H_{2} + H_{3},
\end{equation}
where:
\begin{equation}\label{H1}
H_{1} = -\delta + p_z ,
\end{equation}
\begin{equation}\label{H2}
H_{2} = \frac{(p_x - a_y)^2}{2(1+\delta)},
\end{equation}
\begin{equation}\label{H3}
H_{3} = \frac{(p_y - a_y)^2}{2(1+\delta)}.
\end{equation}
so the Lie transformation can now be written as:

\begin{equation}\label{liesplit}
\begin{split}
e^{-\Delta \sigma:\overline{H}:} 
& \approx e^{-\Delta \sigma:H_1+H_2+H_3:} \\
& \approx e^{-\frac{\Delta \sigma}{2}:H_1:}e^{-\frac{\Delta \sigma}{2}:H_2:}e^{\Delta \sigma:H_3:}e^{-\frac{\Delta \sigma}{2}:H_2:}e^{-\frac{\Delta \sigma}{2}:H_1:}.
\end{split}
\end{equation}

It is worth noting that $H_1$ is exactly solvable while $H_2$ and $H_3$ are not integrable due to the mixed coordinates and their conjugate canonical momenta.Taking $e^{-\Delta \sigma:H_2:}$ as an example, it
can be expressed as a composition of Lie transformations with integrable generators by using the generating function technique. According to 
\begin{equation}\label{liegf}
\begin{aligned}
e^{:f:}e^{:g:}e^{:f:} = e^{:{e^{:f:}}g:}.  
\end{aligned}
\end{equation}
the generating function should be built as:
\begin{equation}\label{liegf_H2}
\begin{aligned}
e^{:I_x:}p_x = p_x - a_x.
\end{aligned}
\end{equation}
so $e^{-\Delta \sigma:H_2:}$ can be written as:

\begin{equation}\label{splitted_H2}
\begin{aligned}
e^{-\Delta \sigma:H_2:} = e^{:I_x:}e^{-\Delta \sigma:\frac{{p_x}^2}{2(1+\delta)}:}e^{-:I_x:},
\end{aligned}
\end{equation}
and the function $I_x$ is given by:
\begin{equation}\label{Ix}
\begin{aligned}
I_x = \int a_x(x,y,z)dx.
\end{aligned}
\end{equation}
Similarly the Lie map  $e^{-\Delta \sigma:H_3:}$ is equivalent to the following form:

\begin{equation}\label{splitted_H3}
\begin{aligned}
e^{-\Delta \sigma:H_3:} = e^{:I_y:}e^{-\Delta \sigma:\frac{{p_y}^2}{2(1+\delta)}:}e^{-:I_y:},
\end{aligned}
\end{equation}
with $I_y$ given by:
\begin{equation}\label{Iy}
\begin{aligned}
I_y = \int a_y(x,y,z)dx.
\end{aligned}
\end{equation}

The key explicit formulae for Lie transformation used in Wu-Forest-Robin integrator are listed here:

\begin{equation}\label{Ix_px}
\begin{aligned}
 e^{\pm:I_x:}p_x = p_x \mp a_x.
\end{aligned}
\end{equation}

\begin{equation}\label{Ix_py}
\begin{aligned}
 e^{\pm:I_x:}p_y = p_x \mp \int \frac{\partial a_x}{\partial y}dx,
\end{aligned}
\end{equation}

\begin{equation}\label{Iy_py}
\begin{aligned}
 e^{\pm:I_y:}p_y = p_y \mp a_y,
\end{aligned}
\end{equation}

\begin{equation}\label{Iy_px}
\begin{aligned}
 e^{\pm:I_y:}p_x = p_y \mp \int \frac{\partial a_y}{\partial x}dy,
\end{aligned}
\end{equation}

\begin{equation}\label{H2M_x}
\begin{aligned}
e^{-\frac{\Delta \sigma}{2}:\frac{{p_x}^2}{2(1+\delta)}:}x = x +  \frac{p_x}{(1+\delta) }\frac{\Delta \sigma}{2},
\end{aligned}
\end{equation}

\begin{equation}\label{H3M_y}
\begin{aligned}
e^{-:\frac{{p_y}^2}{2(1+\delta)}:}y = y +  \frac{p_y}{(1+\delta) }\Delta  \sigma.
\end{aligned}
\end{equation}

Essentially, the transformation in Eq.~\eqref{H-eq1} represents 11 successive transformations and is equivalent to a 'drift-kick-drift-kick-drift-kick-drift-kick-drift-kick-drift' approximation. i.e. $e^{\pm:I_x:}p_x$ corresponds to a \emph{kick}, and $e^{-\frac{\Delta \sigma}{2}:\frac{{p_x}^2}{2(1+\delta)}:}x$ to a drift. However, there is no unique form for the transformation in Eq.~\eqref{H1H2H3}. It depends on the magnetic field gauges used in the Hamiltonian and how the Hamiltonian is split.

\subsection{Analytical generating function method} \label{sectionAGF}

In Ref.~\cite{bahrdt_wuestefeld_2011}, Bahrdt and Wüstefeld developed a symplectic method which derives the dynamic variables stepwise from the integral of Hamiltonian with respect longitudinal coordinate z. It is realized by building a mixed-variable generating function (GF) of the third kind which satisfies the Hamiltonian-Jacobian equation. The canonical transformation between the initial dynamical variables $(x,p_x,y,p_y)$ and final ones $(x_f,p_{xf},y_f,p_{yf})$ uses a relation of the form in~\crefrange{HJE}{GFpy}.

\begin{equation}\label{HJE}
\begin{aligned}
\frac{\partial F_3(x,p_{xf},y,p_{yf})}{\partial z}= -H,
\end{aligned}
\end{equation}

\begin{equation}\label{GFxf}
\begin{aligned}
  x_f = -\frac{\partial F_3(x,p_{xf},y,p_{yf})}{\partial p_{xf}},
\end{aligned}
\end{equation}

\begin{equation}\label{GFyf}
\begin{aligned}
y_f = -\frac{\partial F_3(x,p_{xf},y,p_{yf})}{\partial p_{yf}},
\end{aligned}
\end{equation}
\begin{equation}\label{GFpxi}
\begin{aligned}
p_{x} = -\frac{\partial F_3(x,p_{xf},y,p_{yf})}{\partial x},
\end{aligned}
\end{equation}
\begin{equation}\label{GFpy}
\begin{aligned}
p_{y} = -\frac{\partial F_3(x,p_{xf},y,p_{yf})}{\partial y}.
\end{aligned}
\end{equation}

To construct the GF from Hamiltonian according to~\crefrange{HJE}{GFpy}, $F_3$ is expressed as:
\begin{equation}\label{F3}
\begin{aligned}
F_3  & = -\int H dz \\
     & = -\int [-1 + \frac{(p_x - a_x)^2}{2} + \frac{(p_y - a_y)^2}{2} - a_z]dz \\
     &+ \widetilde{F}(x,p_{xf},y,p_{yf}).
\end{aligned}
\end{equation}
so the Hamiltonian-Jacobian equation has the new form:
\begin{equation}\label{HJE2}
\begin{aligned}
H(x,p_{xf},y,p_{yf}) + \frac{\partial F_3(x,p_{xf},y,p_{yf})}{\partial z} = 0.
\end{aligned}
\end{equation}

Choosing a series of Taylor expansion to represent the GF:
\begin{equation}\label{F3taylor}
\begin{aligned}
F_3 = \sum_{ijk} f_{ijk}p_{xf}^i p_{yf}^j {x_3}^k.
\end{aligned}
\end{equation}
in which $f_{ijk}$ coefficients are functions of position variables $x$, $y$ and $z$, and the expansion order is given by $i+j+k$. Especially, $x_3$ is an order counting number and will be replaced with 1 in the end.

The expansion of the GF can be factorized into field-independent terms and field-dependent terms. The field-independent terms can be derived directly by applying ~\crefrange{GFxf}{GFpy} to a drift section, the other four terms with coefficients $f_{001},f_{002},f_{011},f_{101}$ are added as the field-dependent terms up to the second order. Finally the GF is constructed as follows:
\begin{equation}\label{F3const}
\begin{aligned}
F_3= z_f-(p_{xf}x+p_{yf}y)-\frac{({p_{xf}}^2+{p_{yf}}^2)z_f}{2}\\+f_{101}p_{x_f}x_3+f_{011}p_{y_f}x_3+f_{002}{x_3}^2+f_{001}{x_3},
\end{aligned}
\end{equation}
Inserting $F_3$ to Eq.~\eqref{HJE2} and abandoning the terms higher than second order, the expanded form of the HJE becomes:

\begin{equation}\label{NEWHJE}
\begin{aligned}
&-1 + (-p_{xf}+f_{001x}x_3+A_x x_3)^2/2 \\
&+ (-p_{yf}+f_{001y}x_3 + A_y x_3)^2/2 - A_z x_3 \\
&+ 1 - (p_{xf}^2 + p_{yf}^2)/2 + f_{101z}p_{xf}x_3 \\
&+ f_{011z}p_{yf}x_3 + f_{002z}x_3^2 + f_{001z}x_3=0.
\end{aligned}
\end{equation}
in which the partial derivatives of $f_{ijk}$ to $x$, $y$ and $z$ are expressed as $f_{ijkx}$,$f_{ijky}$ or $f_{ijkz}$. The coefficients are solved by eliminating the terms with the same order:
\begin{equation}\label{f001z}
\begin{aligned}
  f_{001z} = A_{z}, 
\end{aligned}
\end{equation}
\begin{equation}\label{f011z}
\begin{aligned}
  f_{011z} = f_{001y} + A_y, 
\end{aligned}
\end{equation}
\begin{equation}\label{f101z}
\begin{aligned}
  &f_{101z} = f_{001x} + A_x, 
\end{aligned}
\end{equation}
\begin{equation}\label{f002z}
\begin{aligned}
  f_{002z} = -\frac{1}{2}(f_{001x} + A_x)^2 -\frac{1}{2} (f_{001y} + A_y)^2.  
\end{aligned}
\end{equation}
therefore the analytical expressions of $f_{001},f_{002},f_{011},f_{101}$ can be obtained as follows: 
\begin{equation}\label{f001}
\begin{aligned}
  f_{001} = \int A_{z} dz,
\end{aligned}
\end{equation}
\begin{equation}\label{f002}
\begin{aligned}
f_{002} =& -\frac{1}{2} \int [ (A_x + \int \frac {\partial A_z} {\partial x}dz')^2 \\
            &\quad \quad+ (A_y + \int \frac {\partial A_z} {\partial y}dz')^2] dz,
\end{aligned}
\end{equation}
\begin{equation}\label{f011}
\begin{aligned}
f_{011} = \int (A_y + \int \frac {\partial A_z} {\partial y}dz')dz,
\end{aligned}
\end{equation}
\begin{equation}\label{f101}
\begin{aligned}
f_{101} = \int (A_x + \int \frac {\partial A_z} {\partial x}dz')dz. 
\end{aligned}
\end{equation}
now inserting $F_3$ again to~\crefrange{GFxf}{GFpy}, finally the explicit transfer map is given by:
\begin{equation}\label{AGFxf}
\begin{aligned}
x_f = x - f_{101} + p_{xf}z_f,
\end{aligned}
\end{equation}

\begin{equation}\label{AGFpxf}
\begin{aligned}
p_{xf} =& [(1-f_{011y})(p_x + f_{002x} + f_{001x}) \\ 
            &+ f_{011x}(p_y + f_{002y} + f_{001y})]/ {p_n},
\end{aligned}
\end{equation}

\begin{equation}\label{AGFyf}
\begin{aligned}
y_f = y -f_{011} + p_{yf}z_f,
\end{aligned}
\end{equation}

\begin{equation}\label{AGFpyf}
\begin{aligned}
p_{yf} =& [(1 - f_{101x})(p_y + f_{002y} + f_{001y}) \\
  &+ f_{101y}(p_x + f_{002x} + f_{001x})]/p_n,
\end{aligned}
\end{equation}
in which $p_n$ is given by:
\begin{equation}\label{AGFpn}
\begin{aligned}
p_n = (1-f_{011y})(1-f_{101x}) - f_{011x}f_{101y}.
\end{aligned}
\end{equation}

It is reported in Ref.~\cite{bahrdt_wuestefeld_2011} that a higher order expansion with only  the $x_3$ variable will increase the accuracy of the approximation and can still be solved in a similar way. A higher order expansion with momenta needs be solved by the Newton-Raphson method. Either type of  higher order terms significantly increases the computation cost, therefore the expansion used in this paper is only up to the second order.

It is worth pointing out that the analytical generating function method allows integration through a whole period of a insertion device in one single step on the condition of applying proper analytical representation, therefore many terms ($sin$ terms with $z$) of Fourier decomposition vanish upon integration~\cite{bahrdt_wuestefeld_2011}, and this dramatically speeds up the computation. However, the expressions in Eq.~\eqref{Mby} and Eq.~\eqref{MbyAPPLEII} are not optimized for fast one-period integration in a single step with analytical generating function method. In addition, one-period length of the RW has been proved in practice to be too large for one integration step and leads to nonphysical results.

\section{symplectic tracking via monomial maps}\label{trackingmono}
The integration methods above need to calculate the derivative of the magnetic field for each integration step, therefore they are very time-consuming for multi-turn particle tracking. In addition, it is a huge effort to obtain an accurate analytical expression of the magnetic field.

In Ref.~\cite{li_huang_2015}, a practical tracking approach was proposed without knowing the analytical expression of the magnetic field. The authors first extract the Taylor map of an arbitrary field from one-pass multi-tracking, and then convert the Taylor map into a Lie map, which yields to a train of monomial maps by factorization. Most importantly each monomial map has explicit solutions. Inspired by this approach, the mononial map method is employed for tracking through the RW. However, we skip the tedious steps from Taylor map to the monomial map, and fit the coefficients of monomial map directly to the one-pass multi-particle tracking. Moreover, the one-pass multi-particle tracking doesn't have to be symplectic, because the monomial map is symplectic by nature.

To simplify the fitting, 4D monomial maps without $l$ and $\delta$ is used in this paper. The transfer map $\mathcal{M}$ applied to the Robinson wiggler composes of a series of monomial maps up to the $9^{th}$ order as follows:
\begin{equation}\label{monoG29}
\begin{aligned}
 \mathcal{M} = e^{:G_2:}e^{:G_3:}e^{:G_4:}e^{:G_5:}e^{:G_6:}e^{:G_7:}e^{:G_8:}e^{:G_9:}.
\end{aligned}
\end{equation}
in which the second order map is expressed as:
\begin{equation}\label{monoG2}
\begin{aligned}
 e^{:G_2:}=~& e^{a_{2000}:x^2:}e^{a_{1100}:x p_x:}e^{a_{0200}:p_x^2:}e^{a_{1010}:x y:}e^{a_{0110}:p_x y:}\\
 & e^{a_{0020}:y^2:}e^{a_{1001}:x p_y:}e^{a_{0101}:p_x p_y:}e^{a_{0011}:y p_y:}e^{a_{0002}: p_y^2:}.
\end{aligned}
\end{equation}

Basde on the 2D formulae in Ref.~\cite{chao2002}, explicit solutions of a 4D monomial map are given by:

\begin{equation}\label{monox}
\begin{aligned}
     & e^{a:x^k p_x^l y^mp_y^n:}x \\ & =\left\{
                \begin{array}{ll}
                  x [1+a(k-l)x^{k-1}p_x^{l-1}y^mp_y^n]^{l/(l-k)},& \text{if } k\neq l\\
                  x e^{-a k x^{k-1}p_x^{k-1}y^mp_y^n},&             \text{if } k= l\\
                \end{array}
              \right. 
\end{aligned}
\end{equation}

\begin{equation}\label{monopx}
\begin{aligned}
    & e^{a:x^k p_x^l y^m p_y^n:}p_x\\
    &=\left\{
                \begin{array}{ll}
                  p_x [1+a (k-l) x^{k-1}p_x^{l-1}y^mp_y^n]^{k/(k-l)},& \text{if } k\neq l\\
                  p_x e^{a k x^{k-1}p_x^{k-1}y^mp_y^n},&             \text{if } k= l\\
                \end{array}
              \right.
\end{aligned}
\end{equation}

\begin{equation}\label{monoy}
\begin{aligned}
    &e^{a:x^k p_x^l y^mp_y^n:}y\\
    &=\left\{
                \begin{array}{ll}
                  y [1 + a(m-n)y^{m-1}p_y^{n-1}x^k p_x^l]^{n/(n-m)},& \text{if } m\neq n\\
                  y e^{-a m y^{m-1}p_y^{m-1}x^k p_x^l},&             \text{if } m= n\\
                \end{array}
              \right.
\end{aligned}
\end{equation}

\begin{equation}\label{monopy}
\begin{aligned}
    &e^{a:x^k p_x^l y^mp_y^n:}p_y \\
    &=\left\{
                \begin{array}{ll}
                  p_y [1 + a(m-n)y^{m-1}p_y^{n-1}x^k p_x^l]^{m/(m-n)},& \text{if } m\neq n\\
                  p_y e^{a m y^{m-1}p_y^{m-1}x^k p_x^l},&             \text{if } m= n\\
                \end{array}
              \right.
\end{aligned}
\end{equation}

Based on~\cref{monox,monopx,monoy,monopy}, $e^{:G_2:}$ can be reconstructed from the linear transfer matrix, which is calculated numerically from the field map of the RW. In this step, the symplectic error of the numerical transfer matrix is rounded off due to the intrinsic symplecticity of the monomial map. The coefficients of higher order terms are fitted to the input and output of one-pass multi-particle tracking.

\begin{figure}[htb]
   \centering
   \includegraphics*[width=0.48\textwidth]{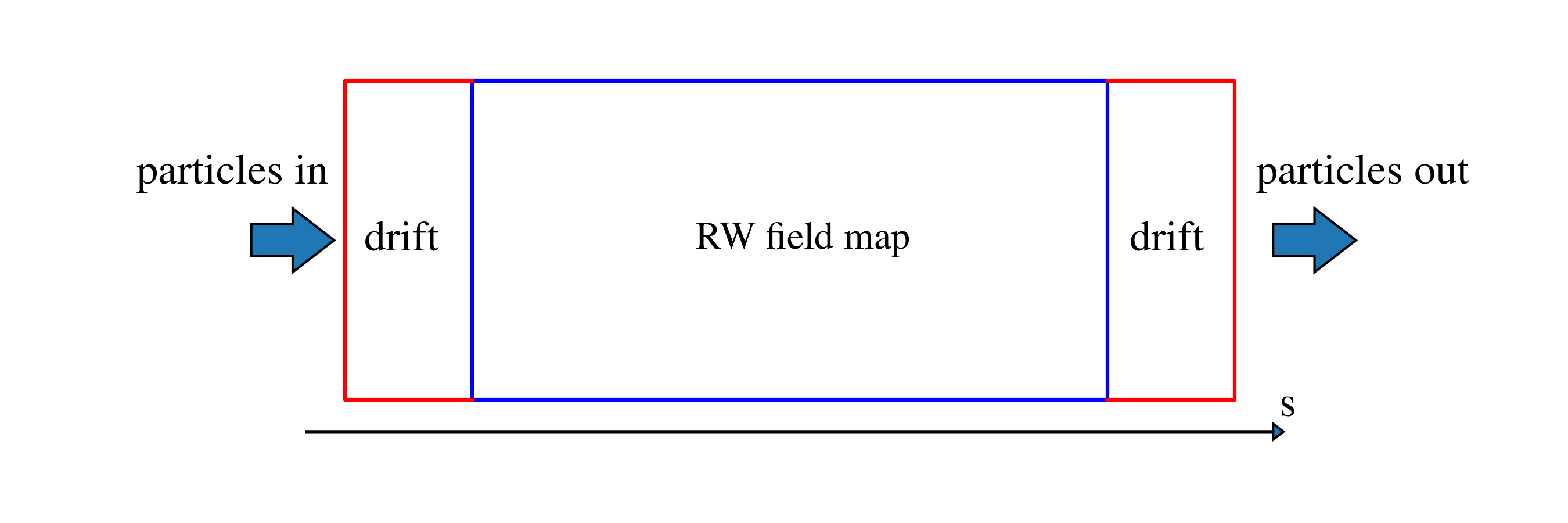}
     \caption{The method used for one-pass multi-particle tracking without the analytical representation of the magnetic field of the RW}
   \label{monofit}
\end{figure}

The field map takes up 2.3~m longitudinally, and there is a small residual magnetic field at the entrance and exit of the field map. Without knowing the analytical expression of the field, the tracking can be done with ordinary Runge-Kutta methods, using ($x$,$x'$,$y$,$y'$) to describe the particle motion. It introduces errors when converting $x'$, $y'$ to $p_x$, $p_y$ for the input and output particles according to~\crefrange{cordsconversion_xp}{cordsconversion_yp}, if the residual field is simply ignored. As shown Fig~\ref{monofit}, two 5~mm drifts are added before and after the field map, thereby saving the trouble of unknown $a_x$, $a_y$ at the entrance and exit of the field map. However, as the one-pass multi-particle tracking is done with the integrators in section \ref{integratorswithAfield}, the conversion process described above is not needed.

The monomial map method has the advantage of treating the whole RW as a black box. Moreover, it can be universally applied to tracking through elements with arbitrary field, and need not to be limited to the cases where the analytical field expressions are unknown or hard to obtain.

The ranges of dynamic variables of input particles are highly relevant for the accuracy of the monomial map. Here the concept of acceptance from momentum acceptance in nonlinear beam dynamics simulations is borrowed for illustration. The acceptance of $x$ and $y$ are determined by the vacuum chamber size of the RW. Taking $p_x$ as an example, the absolute value of $p_x$ is increased step by step with both positive and negative signs, while $x$,$y$, $p_y$ are set as 0, until the particle is lost in one-pass tracking, so that the acceptance of $p_x$ is obtained. Each dynamic variable of the input particles should be sampled uniformly in the range bounded by the acceptance. If the dynamic variable is only sampled in the paraxial region, the monomial map cannot describe the motion of the particles with large excursion or large momenta. The order of the monomial map is another crucial factor for the accuracy, and in practice it is increased until convergence. In this paper monomial map up to $9^{th}$ order is a compromise between accuracy and computation efficiency.

Fitting the coefficients of the monomial map is challenging work, however, it is much easier than reconstructing the magnetic field analytically. The fitting routine here is based on gradient descent method, with the help of CUDA GPU acceleration and automatic differentiation in Pytorch.

\section{comparison}

\begin{figure*}[tbh!]
  \centering
  \includegraphics*[width=0.245\textwidth]{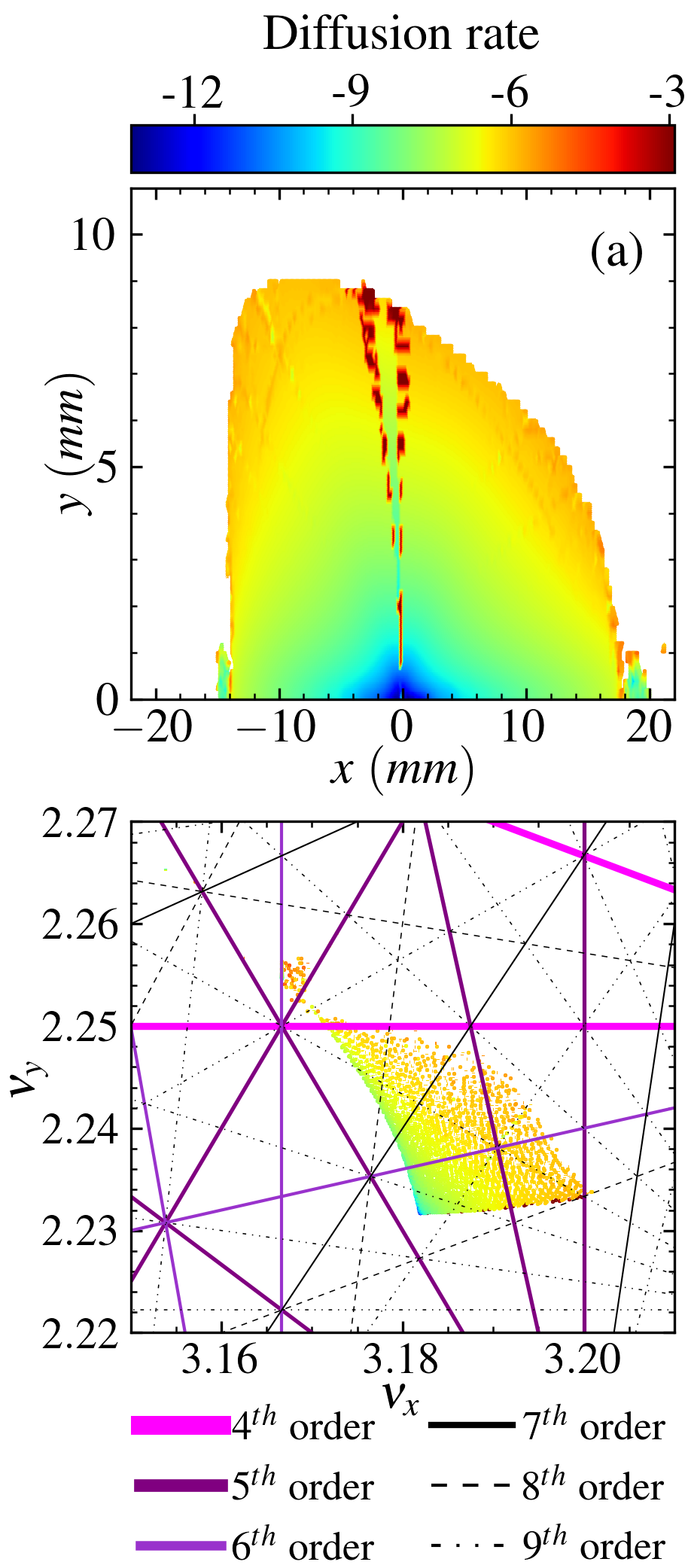}
  \includegraphics*[width=0.245\textwidth]{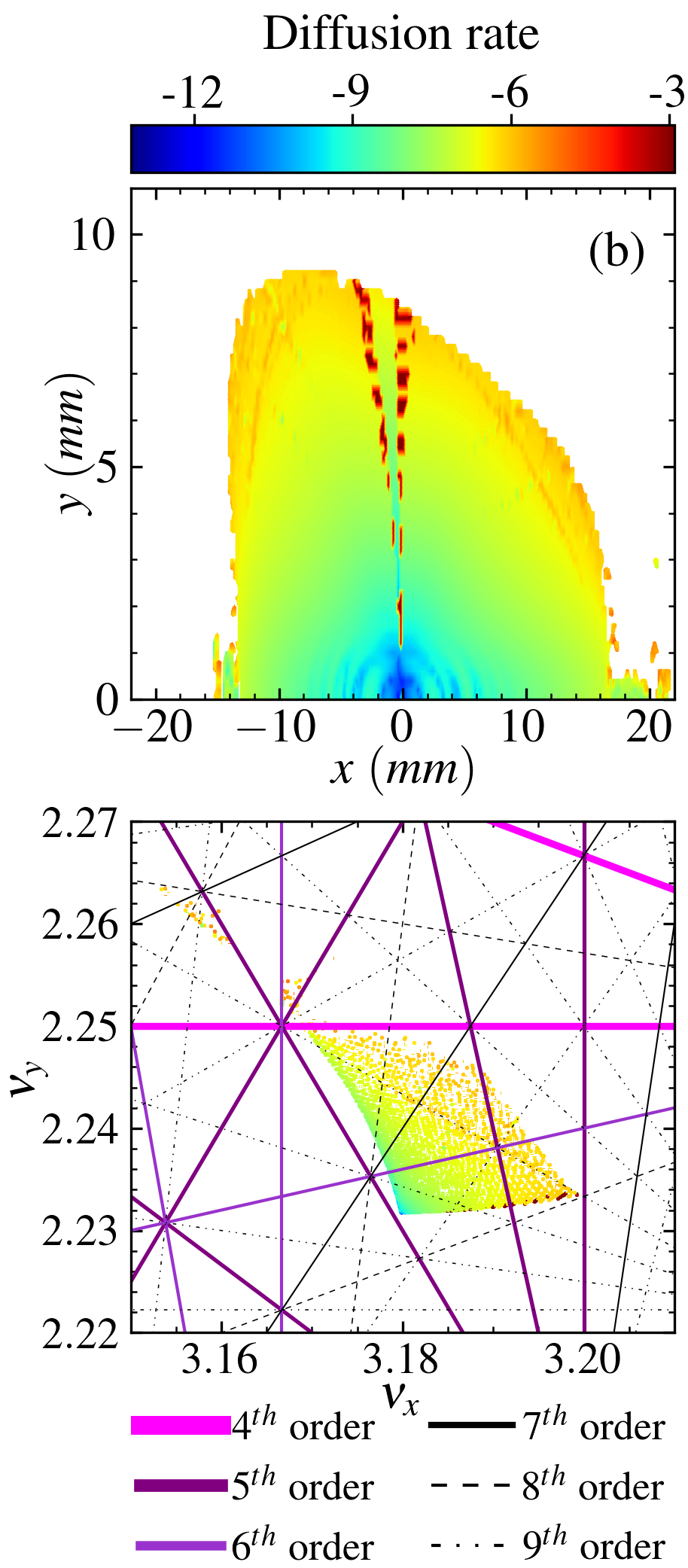}
  \includegraphics*[width=0.245\textwidth]{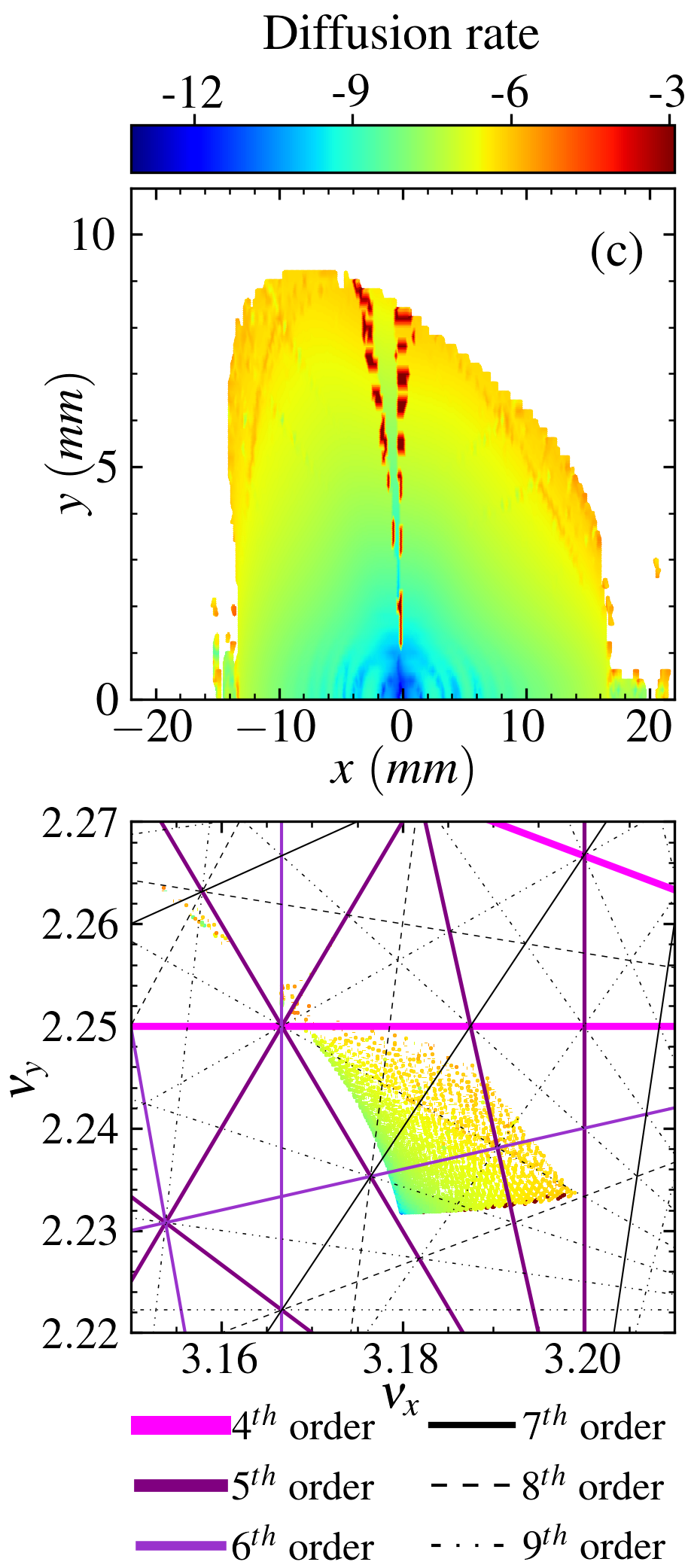}
  \includegraphics*[width=0.245\textwidth]{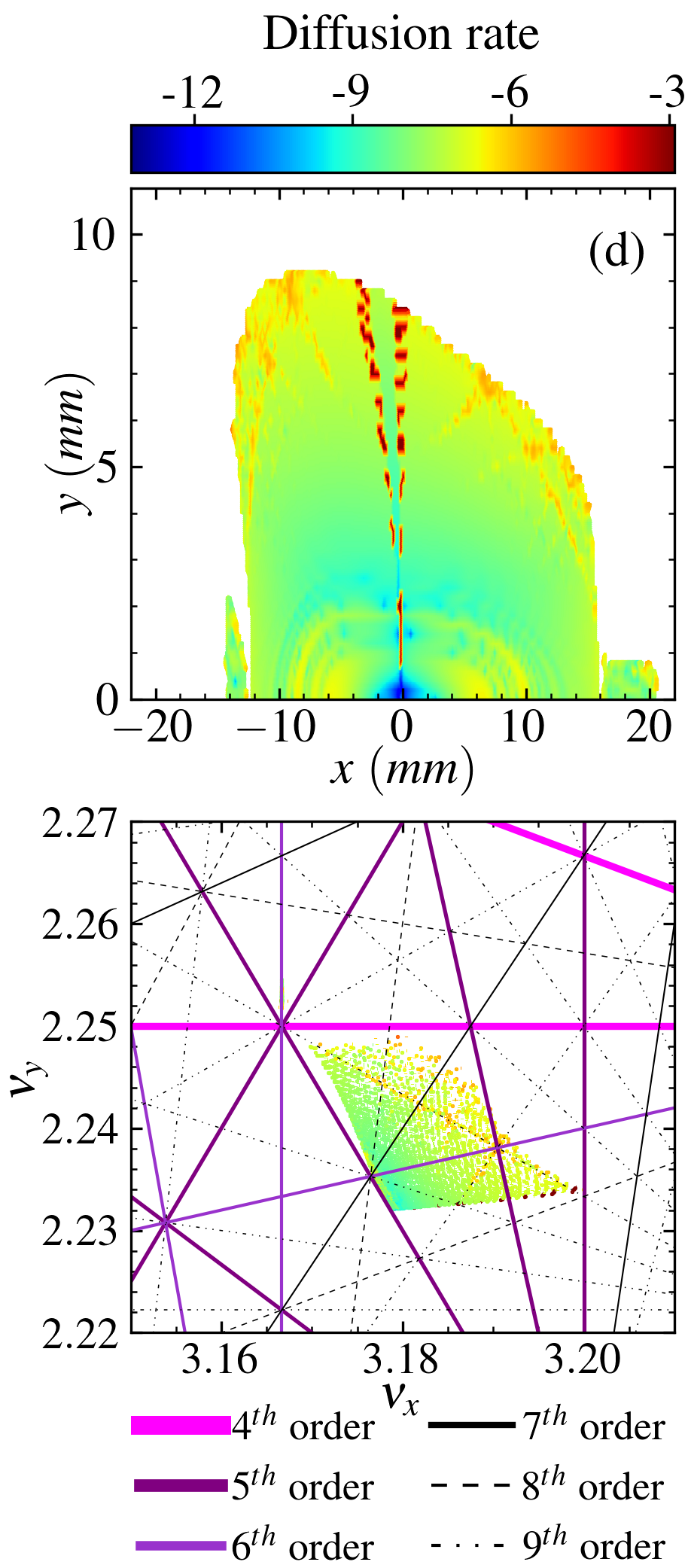}
  \caption{Dynamic aperture with diffusion rate based on (a) implicit Runge-Kutta integrator (b) Wu-Forest-Robin integrator (c) analytical generating function method (d) monomial map }
  \label{fmacomp}
\end{figure*}

 The nonlinear distortion of the RW to the beam motion is measured with Frequency Map Analysis (FMA), realized with ELEGANT~\cite{borland_2000}. 
 Although there is no module in ELEGANT which can represent the Robinson wiggler directly, the SCRIPT element provides an interface to use customized integrators tracking through the Robinson wiggler and to make use of the powerful analysis tools in ELEGANT. The results based on four integrators above are plotted in Fig.~\ref{fmacomp}.
 
In the FMA plots, the Dynamic Aperture (DA) is given by the boundary of survival particles after tracking 1024 turns. The diffusion rate $d_r$ is defined as follows:
\begin{equation}\label{eq:diff_rate}
\begin{aligned}
d_r = \log_{10} \frac { \sqrt{{\Delta \nu_x}^2+{\Delta \nu_y}^2} }{N}.
\end{aligned}
\end{equation}
in which $\Delta \nu_x$ and $\Delta \nu_y$ are the differences in horizontal and vertical tunes from the first and second half turns of the tracking, and N is the number of tracking turns.

As shown in Fig.~\ref{fmacomp} (a), (b) and (c), the integrators based on an analytic field expression give very similar results in terms of DA and the tune footprint. The differences of the results between the Wu-Forest-Robin integrator and the analytical generating function method are negligible. However, the implicit Runge-Kutta integrator uses an exact Hamiltonian without approximation and yields to larger diffusion rates for large-amplitude particles. By contrast the Wu-Forest-Robin integrator and the analytic generating function method make use of the approximated Hamiltonian to get rid of the mixed terms of coordinates and their momenta, so the nonlinear coupling effect has been weakened artificially which leads to smaller diffusion rates for large-amplitude particles.

The monomial map gives a very similar DA and slightly different tune footprint. The reason is that the settings of the quadrupoles and sextupoles for tracking are preliminarily optimized based on the numerical field map. The analytical field representation has a discrepancy with the numerical field map, the residual of the Fourier decomposition cause a small but non-negligible distortions on the orbit and $\beta$ functions. In contrast, the monomial map rounds off the orbit distortion by abandoning the first order terms. The second-order terms of the monomial map are directly reconstructed from the linear transfer matrix of the field map, therefore there is no distortion on $\beta$ functions. Nevertheless, the discrepancies among the four integrators on DA and tune footprint are very small.  

Tracking with the implicit Runge-Kutta integrator is the most time consuming case and takes ~20 times longer than that with monomial maps. The Wu-Forest-Robin integrator takes a similar time as the analytical generating function method, but still ~6 times longer than the monomial map approach. As pointed out in section \ref{sectionAGF}, the feature of the analytical generating function method, which makes Fourier terms vanish by integrating over one period in one step, can make the integration much faster compared to Wu-Forest-Robin integrator with the same step size. However, the suitable objects are limited to short-period undultors described with an optimized analytical representation.

\section{Summary}
The analytical representation of magnetic field in the Robinson wiggler has been established based on modified Halbach expansions and shows very good accuracy when describing the numerical field map. It is generally applicable for insertion devices.

Three integrators based on analytical form of the Hamiltonian are introduced to realize symplectic tracking.  These integration methods are in general very time-consuming for multi-turn tracking. As an alternative approach, the monomial map method shows the advantages of faster speed and saving the trouble of Fourier decomposition. However, the coefficients of the monomial map must be fitted carefully and the orders should be properly chosen. 

The nonlinear dynamics study is performed with ELEGANT and customized integrators. The FMA results based on the implicit Runge-Kutta integrator, Wu-Forest-Robin integrator, analytical generating function method and momomial map are consistent and cross validate one another.

There are other symplectic tracking methods not included in this paper, such as the widely used kick map method~\cite{Elleaume} and  numerical generating function method~\cite{Scheer1992,Scheer2008,Li2019}. We consider that the kick map method essentially treats one period of the insertion device as a thin element, which does not conform with the idea of modeling insertion devices as 3D-field elements. Furthermore, it is found that the numerical generating function method in practice can only describe the motion of particles accurately in paraxial region or in weak magnetic fields. Those two methods can be used for symplectic tracking with much faster speed, however the results should be benchmarked with the integrators used in this paper.

\section{Acknowledgement}
The authors would like to thank Andreas Jankowiak (HZB) and Mathias Richter (PTB) for ongoing support, and Tobias Tydecks (formerly at HZB) for his excellent work on the RW. We thank Ji-Gwang Hwang (HZB) for the fruitful discussions on wide topics in beam dynamics, Michael Scheer for the discussions on numerical generating function method, Zhouyu Zhao (USTC/NSRL) for providing the field map of a UE40 undulator, as well as Xiaobiao Huang (SLAC) and Laurent Nadolski (SOLEIL) on modeling insertion devices in AT, especially Yongjun Li (BNL/NSLS II) for his detailed suggestions on tracking techniques. The deepest gratitude goes to Godehard Wüstefeld (HZB) for inspiring  discussions and detailed guidance on the RW.

\appendix

\section{Analytical field representation for APPLE~II undulator}\label{APPLIIappendix}

In contrast to planar undulators (or wigglers) or the RW,  the magnetic field in APPLE~II undulator is in general not symmetric vertically, hence the field representation is modified to Eq.~\eqref{MbyAPPLEII}.  It is important to point out that Equation~\eqref{Mby} or Equation~\eqref{MbyAPPLEII} describes insertion devices with a fixed field. 
%\begin{widetext}
\begin{equation}
\begin{aligned}
B_y =&\sum_{m,n}^{M,N}cos(mk_xx+\theta_{mn})sin(nk_{z}z +\phi_{mn})\\
   & \times(C_{mn}e^{k_{y,mn}y}+S_{mn}e^{-k_{y,mn}y}).
\label{MbyAPPLEII}
\end{aligned}
\end{equation}
%\end{widetext}

\begin{figure*}[tbh!]
  \centering
  \includegraphics*[width=0.4\textwidth]{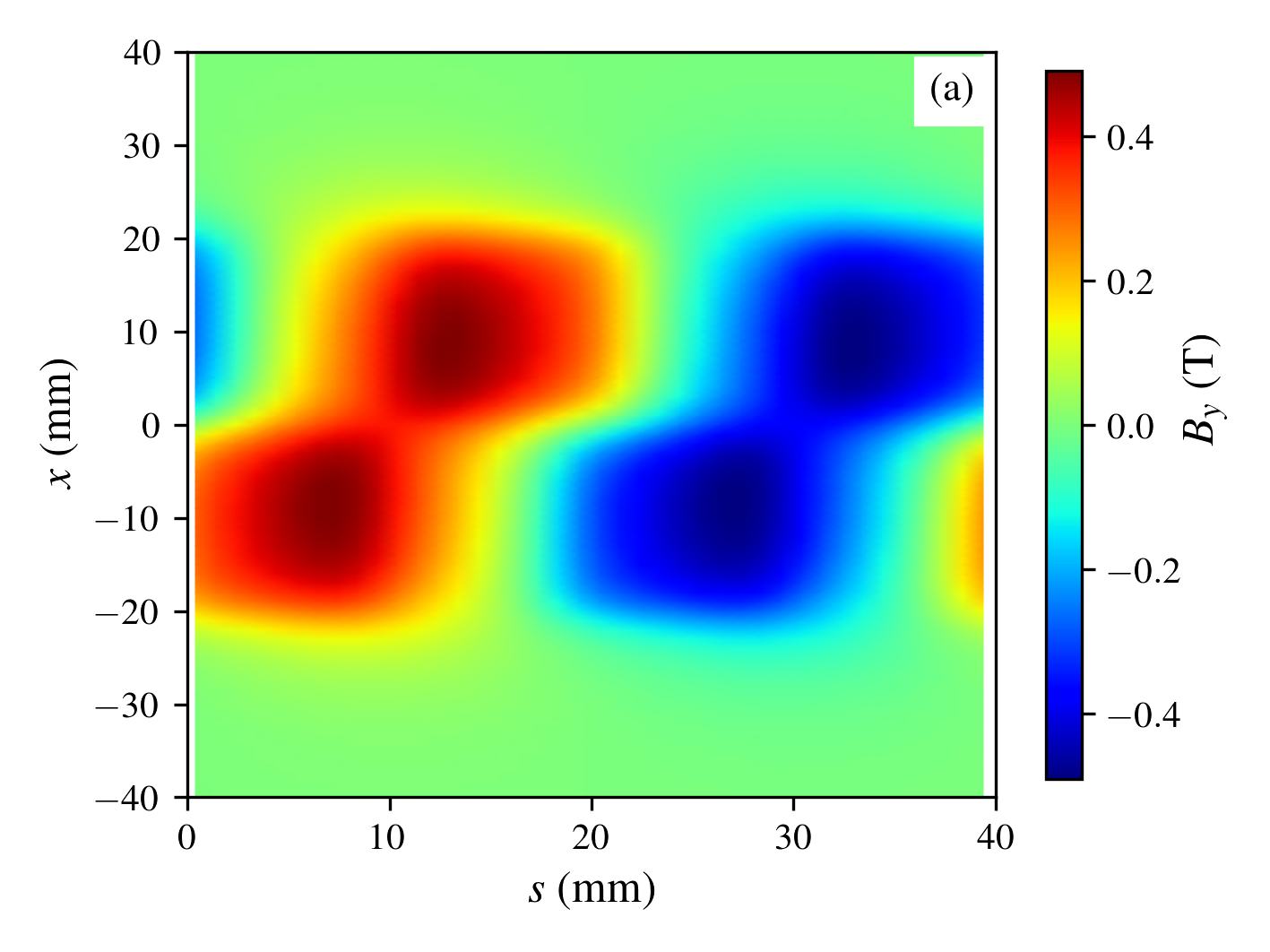}
  \includegraphics*[width=0.4\textwidth]{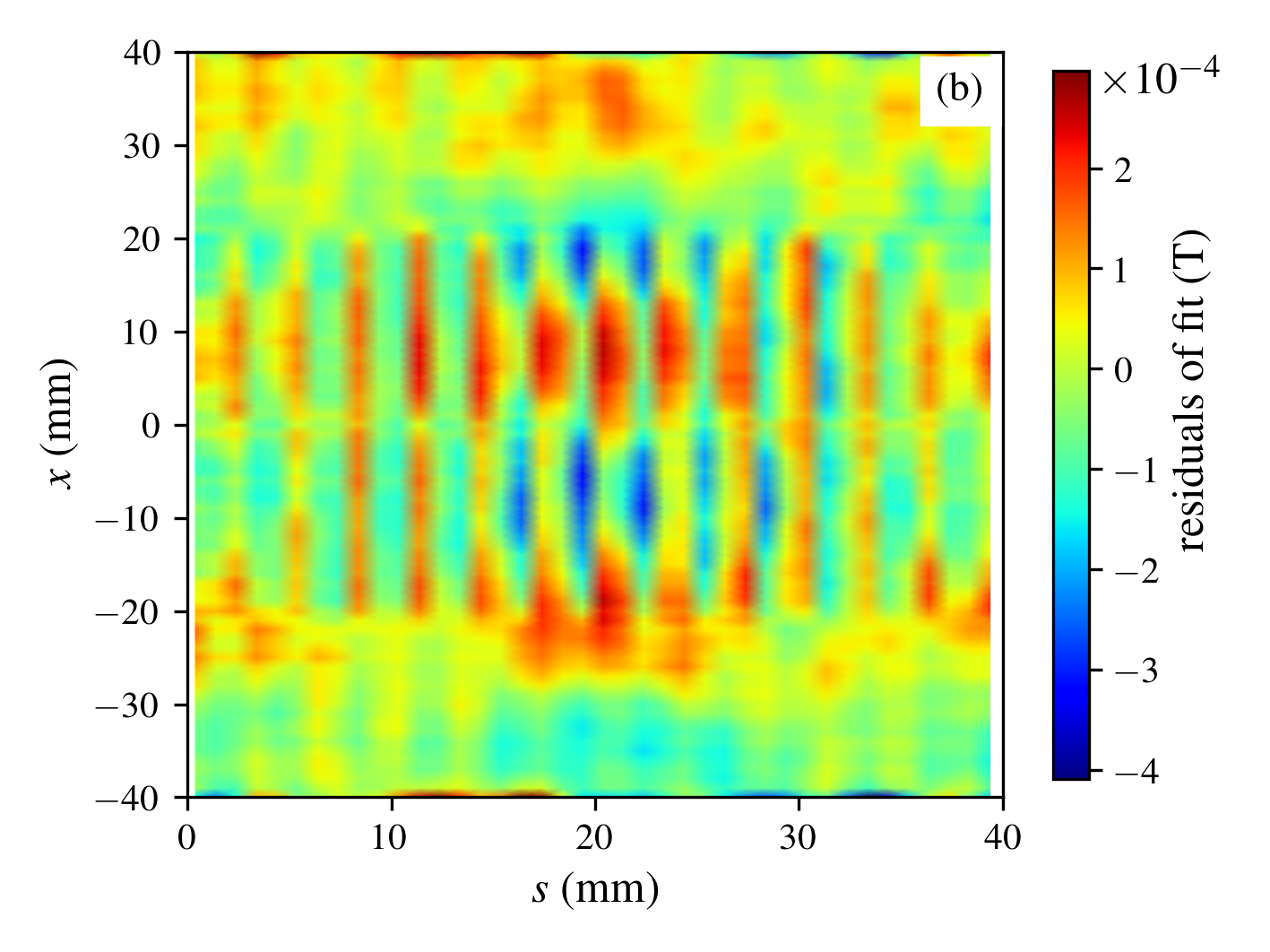}
  \caption{UE40 undulator:(a) vertical magnetic field  (b) fitting residuals   on y = 3~mm plane }
  \label{ue40_fieldmap}
\end{figure*}

We apply the Eq.~\eqref{MbyAPPLEII} to the field map of a UE40 undulator model at USTC/NSRL in circular polarization mode with maximum 12 mm gap~\cite{zhouyuzhao2021}. Without knowing the geometry of the vacuum chamber, the vertical magnetic field on $y = 3 $ mm plane is used to verify the accuracy of Fourier decomposition and plotted in Fig.~\ref{ue40_fieldmap} (a). As shown Fig.~\ref{ue40_fieldmap} (b),  the maximum residual of Fourier decomposition is $\sim$ $4 \times 10^{-4}$~T with $M = 32$ and $N = 32$.

%\begin{figure*}[tb!]
%  \centering  
%  \subcaptioninbox{}{\includegraphics[width=0.48\textwidth]{UE40_3mm.png}}
%  \subcaptioninbox{}{\includegraphics[width=0.48\textwidth]{UE40_3mm_residual.png}}
%  \caption{ UE40 undulator:(a) vertical magnetic field on (y = 3mm) plane (b) fitting residual on (y = %3mm) plane }
%  \label{fig:ue40_fieldmap}
%\end{figure*}

\begin{figure*}
  \centering
  \includegraphics*[width=0.4\textwidth]{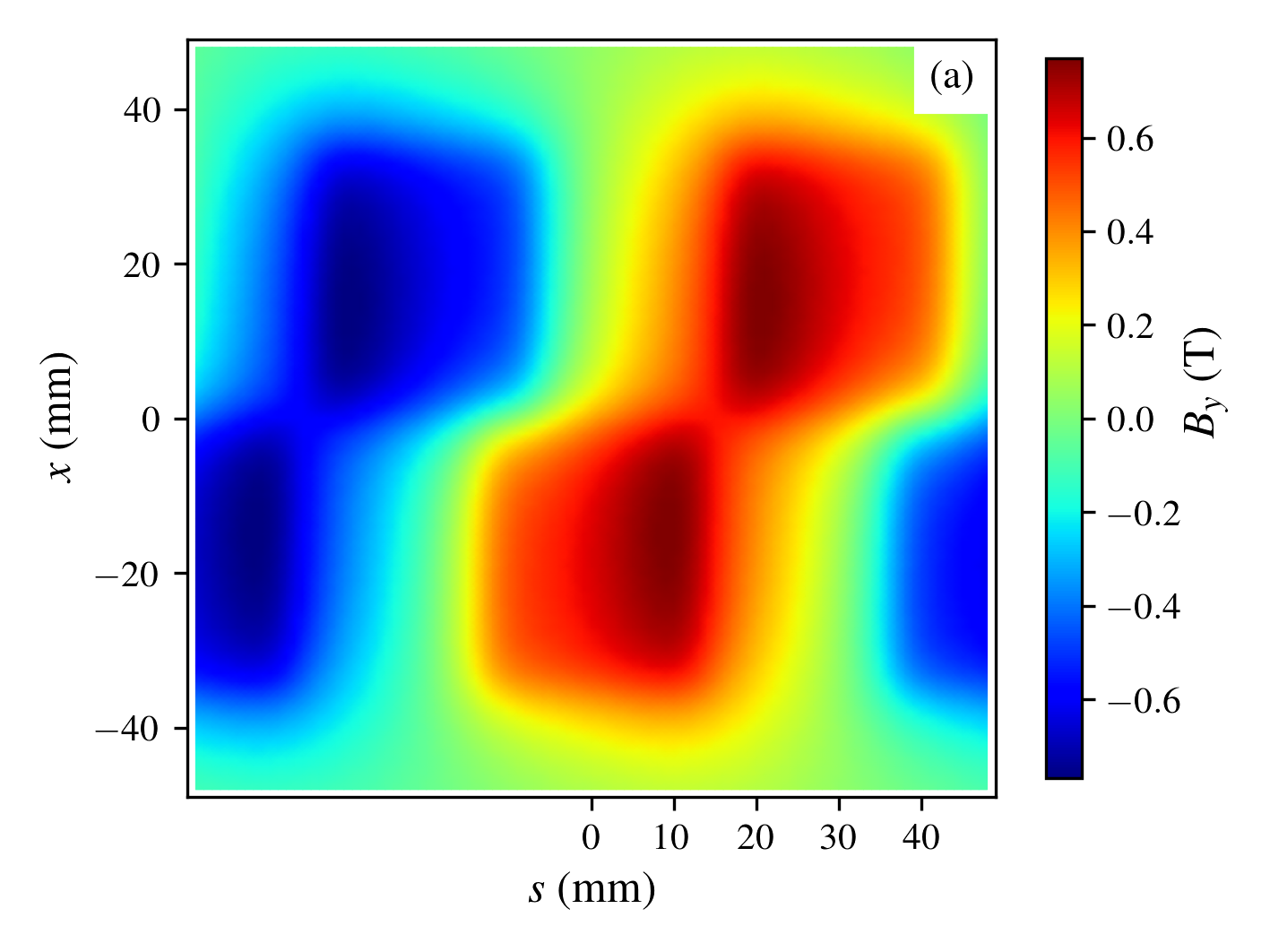}
  \includegraphics*[width=0.4\textwidth]{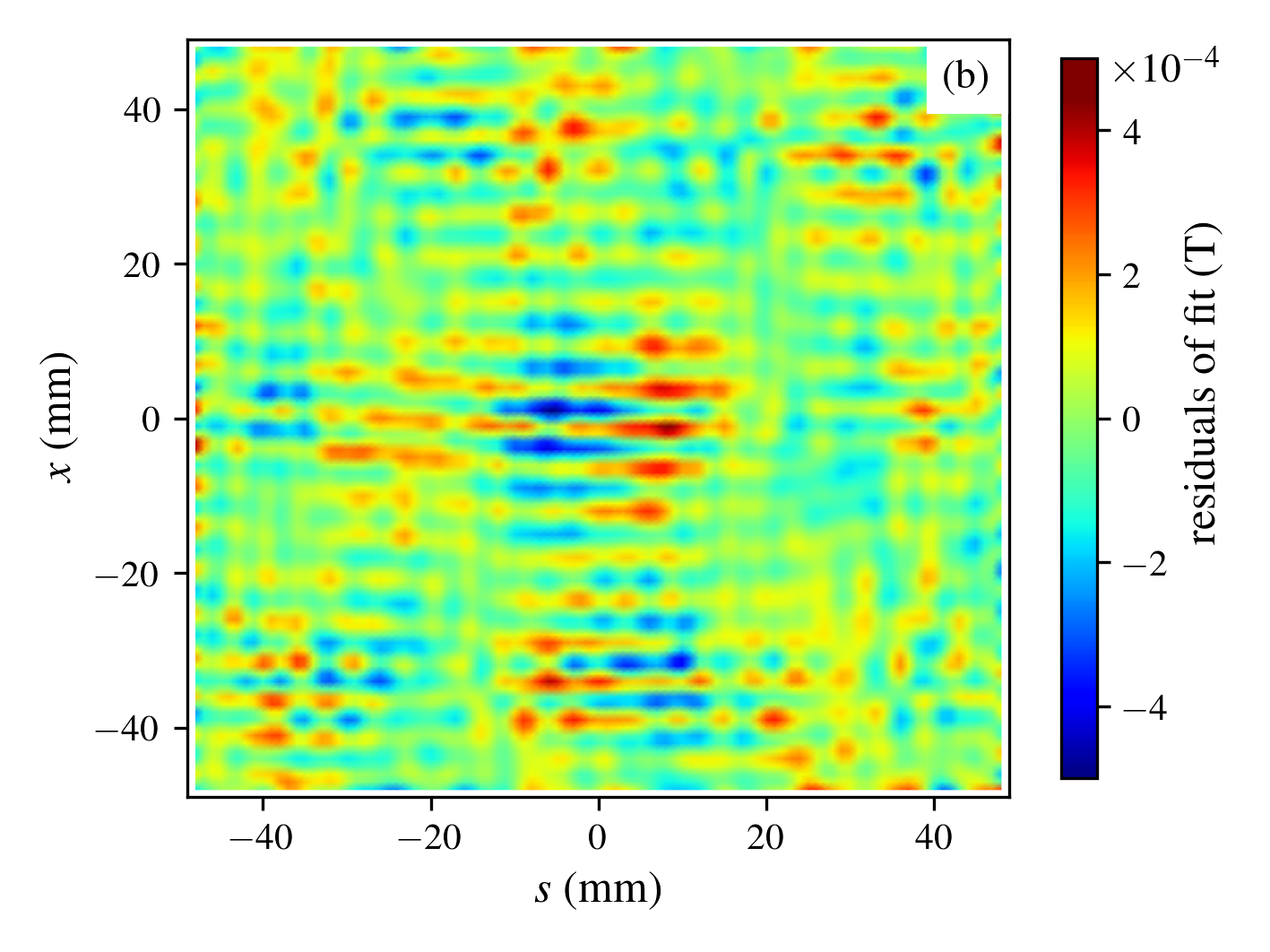}
  \caption{UE100 undulator:(a) vertical magnetic field  (b) fitting residuals   on y = 8~mm plane }
  \label{ue100_fieldmap}
\end{figure*}

If the field dependence on the magnet row movement and gap in an APPLE~II undulator is desired, a dedicated representation of the magnetic field for an APPLE~II is included in Ref.~\cite{bahrdt_wuestefeld_2011}, with which different gaps and polarization modes can be described in one single formula. Here an expression including field dependence on the magnet row movement is given in Eq.~\eqref{APPLEII4S}, in which $\phi_{z1}$, $\phi_{z2}$, $\phi_{z3}$ and $\phi_{z4}$ describe movement of the magnet rows and $\phi_{z1}$ , $\phi_{z2}$ share the same value. 

\begin{widetext}
\begin{equation}
\begin{aligned}
B_y = \sum_{m,n}^{M,N}C_{mn}&(\quad cos(mk_x(x+x_0)+\theta_{mn})sin(nk_{z}z +\phi_{mn} + \phi_{z1} )e^{k_{y,mn}y}\\
    &+cos(mk_x(x-x_0)-\theta_{mn} )sin(nk_{z}z +\phi_{mn} + \phi_{z2})e^{-k_{y,mn}y}\\
    &+cos(mk_x(x+x_0)+\theta_{mn} )sin(nk_{z}z +\phi_{mn} + \phi_{z3})e^{-k_{y,mn}y}\\
   &+cos(mk_x(x-x_0)-\theta_{mn} )sin(nk_{z}z +\phi_{mn}+ \phi_{z4})e^{k_{y,mn}y} \quad).
\label{APPLEII4S}
\end{aligned}
\end{equation}
\end{widetext}

The feasibility of Eq.~\eqref{APPLEII4S} is verified with a UE100 undulator at HZB with maximum 25~mm gap . The vertical magnetic field on $y = 8$~mm plane and the residuals of Fourier decomposition are plotted in Fig.~\ref{ue100_fieldmap}. In practice it is found that Equation~\eqref{APPLEII4S} take fewer harmonics to achieve the same residual level Eq.~\eqref{MbyAPPLEII}  by making use of the transverse symmetric distribution of magnet rows of an APPLE II undulator. With $M = 40$ and $N = 30$, the maximum residual of fit is $\sim$ $5 \times 10^{-4}$~T.

Despite the different expressions, Equation~\eqref{MbyAPPLEII}, Equation~\eqref{APPLEII4S} and the dedicated formulae in Ref.~\cite{bahrdt_wuestefeld_2011} are all linear superpositions of Fourier terms and are essentially the same. Each of them can represent the magnetic field of an APPLE II undulator accurately.

\nocite{*}
\bibliography{apssamp}% Produces the bibliography via BibTeX.

\end{document}